\documentclass[twocolumn,preprintnumbers,superscriptaddress,endnote,nofootinbib,prd]{revtex4-1}

\pdfoutput=1

\usepackage{graphicx}
\usepackage{latexsym}
\usepackage{amsfonts}
\usepackage{amssymb}
\usepackage{amsmath}
\usepackage{slashed}
\usepackage{array}
\usepackage{dcolumn}
\usepackage{feynmp}

\def\lsim{\mathrel{\raise.3ex\hbox{$<$\kern-.75em\lower1ex\hbox{$\sim$}}}}
\def\gsim{\mathrel{\raise.3ex\hbox{$>$\kern-.75em\lower1ex\hbox{$\sim$}}}}

\usepackage{xcolor}
\definecolor{red}{rgb}{1.0, 0, 0}

\newcommand{ \slashchar }[1]{\setbox0=\hbox{$#1$}   
   \dimen0=\wd0                                     
   \setbox1=\hbox{/} \dimen1=\wd1                   
   \ifdim\dimen0>\dimen1                            
      \rlap{\hbox to \dimen0{\hfil/\hfil}}          
      #1                                            
   \else                                            
      \rlap{\hbox to \dimen1{\hfil$#1$\hfil}}       
      /                                             
   \fi}                                             %

\newcommand{\ra}{\rightarrow}
\newcommand{\gev}{\text{GeV}}
\newcommand{\tev}{\text{TeV}}

\newcommand{\fb}{\text{fb}}

\newcommand{\jetsmissing}{nj+\slashchar{E}_T}

\newcommand{\sq}{\tilde{q}}
\newcommand{\Msq}{M_{\tilde{q}}}
\newcommand{\Mg}{M_3}

\begin{document}

\title{Supersoft Supersymmetry is Super-Safe}

\author{Graham D. Kribs}
\affiliation{Department of Physics, University of Oregon,
             Eugene, OR 97403}

\author{Adam Martin}
\affiliation{Theoretical Physics Department, Fermilab, Batavia, IL 60510}
\preprint{FERMILAB-PUB-12-074-T}
\date{\today}

\begin{abstract}

We show that supersymmetric models with a large Dirac gluino mass
can evade much of the jets plus missing energy searches at LHC\@.
Dirac gaugino masses arise from ``supersoft'' operators that lead 
to finite one-loop suppressed contributions to the scalar masses.
A little hierarchy between the Dirac gluino mass $5 \ra 10$ times heavier
than the squark masses is automatic and technically natural, 
in stark contrast to supersymmetric models with Majorana gaugino masses.  
At the LHC, colored sparticle production is suppressed not only
by the absence of gluino pair (or associated) production, 
but also because several of the largest squark pair production channels 
are suppressed or absent.  
We recast the null results from the present jets plus missing energy 
searches at LHC for supersymmetry onto a supersoft supersymmetric 
simplified model (SSSM).  Assuming a massless LSP, we find the 
strongest bounds are:  $748$~GeV from a $2j + \slashchar{E}_T$ 
search at ATLAS ($4.7$~fb$^{-1}$), and $684$~GeV from a combined jets 
plus missing energy search using $\alpha_T$ at CMS ($1.1$~fb$^{-1}$). 
In the absence of a future observation, we estimate the bounds on 
the squark masses to improve only modestly with increased luminosity.  
We also briefly consider the further weakening
in the bounds as the LSP mass is increased.

\end{abstract}

\maketitle

\section{Introduction}
\label{sec:intro}

The parameter space of the minimal supersymmetric standard model 
(MSSM) is significantly constrained by impressive 
searches at the LHC\@.  The strongest limits occur on the mass of 
colored superpartners, well over $1$~TeV in simplified models 
in which the squark or gluino decays to a quark or gluon and a 
(nearly) massless lightest supersymmetric particle (LSP). 
These limits are driven by several search strategies for 
large missing energy and large amounts of hadronic activity, 
which we abbreviate $\jetsmissing$.

The strong constraints from $\jetsmissing$ searches have (re-)motivated
three basic approaches to weak-scale supersymmetry:
\begin{itemize}
\item[1.]  Discard superpartners that are not directly relevant
to electroweak symmetry breaking, including the first and
second generation squarks.  Well known examples are 
more minimal supersymmetry \cite{Dimopoulos:1995mi,Cohen:1996vb} 
and split supersymmetry \cite{ArkaniHamed:2004yi}.  The extent 
to which these approaches successfully retain a light third generation 
with heavy first and second generations have been explored recently 
in several scenarios \cite{Essig:2011qg,Kats:2011qh,Brust:2011tb,Papucci:2011wy,Feng:2011aa,Csaki:2012fh,Larsen:2012rq}.

\item[2.]  Keep superpartners roughly in the sub-TeV region, 
while removing most or all of the \emph{missing energy}, 
thereby rendering $\jetsmissing$ searches moot.  
The classic example is $R$-parity violation 
(for a review, see \cite{Barbier:2004ez}) through the 
baryon number violating $u^c d^c d^c$ term in the superpotential,
which allows the LSP to decay into jets 
(for a recent discussion see 
\cite{Carpenter:2007zz,Csaki:2011ge,Dreiner:2012mn,Allanach:2012vj}). 

\item[3.]  Keep superpartners roughly in the sub-TeV region, 
while removing most or all of the \emph{visible energy}, which 
significantly weakens the effectiveness of $\jetsmissing$ searches.
This approach includes compressed supersymmetry \cite{compressed}, 
stealth supersymmetry \cite{stealth} (which is a hybrid between 
approaches 2 and 3) and others.

\end{itemize}
In this paper, we propose a fourth alternative:
\begin{itemize}
\item[4.]  Keep most superpartners in the sub-TeV region,
while removing much of the \emph{production cross section},
thereby significantly weakening the effectiveness of $\jetsmissing$ searches.
We demonstrate that this alternative allows first and second
generation squarks to be as light as $\sim 650-750$~GeV
with a massless LSP, and potentially even lighter if there
is modest compression by either raising the LSP mass or allowing 
a cascade decay through intermediate mass superpartners.  
The key to this alternative is to assume the gluino acquires 
a large Dirac mass.  
\end{itemize}

Reducing the production cross section of colored superpartners 
``merely'' by raising the 
Majorana gluino mass in the MSSM would seem to be just as sufficient.  
However, the squark masses receive substantial log-enhanced contributions
to their masses through renormalization group evolution.  
This includes the stop masses, which in turn feed into the
Higgs soft masses through the top Yukawa.  
Since the Higgs soft masses directly determine the fine-tuning of
the electroweak symmetry breaking scale, this implies the stops 
as well as the gluino should not far exceed the electroweak scale
without causing excessive unnaturalness.

A heavy Dirac gluino, by contrast, is completely natural
\cite{Fayet:1978qc,Polchinski:1982an,Hall:1990hq,Fox:2002bu}.
Dirac gaugino masses induce one-loop finite contributions to 
squark, slepton and Higgs soft masses from ``supersoft'' 
operators \cite{Fox:2002bu}.
The finiteness implies renormalization group evolution of squark
masses is insensitive to the gaugino masses, preserving the 
little hierarchy $\Mg \simeq (5 \ra 10) \times \Msq$.   
The only price we pay is minimality -- the matter content must be 
extended by three gauge adjoint superfields, one for each gauge group.

\begin{figure*}[t]
\includegraphics[width=137pt]{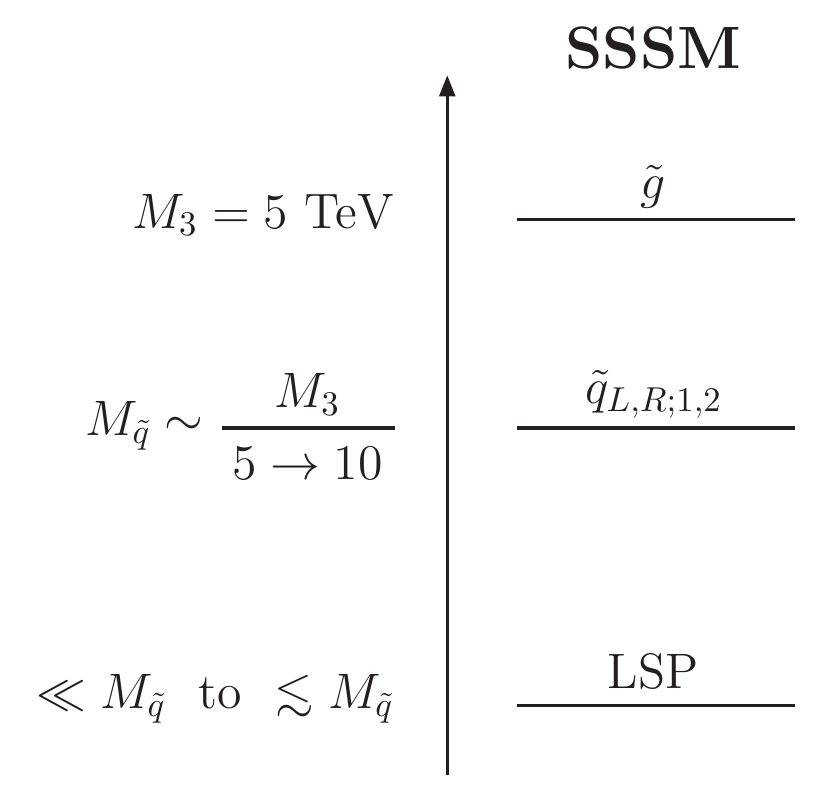}
\includegraphics[width=110pt]{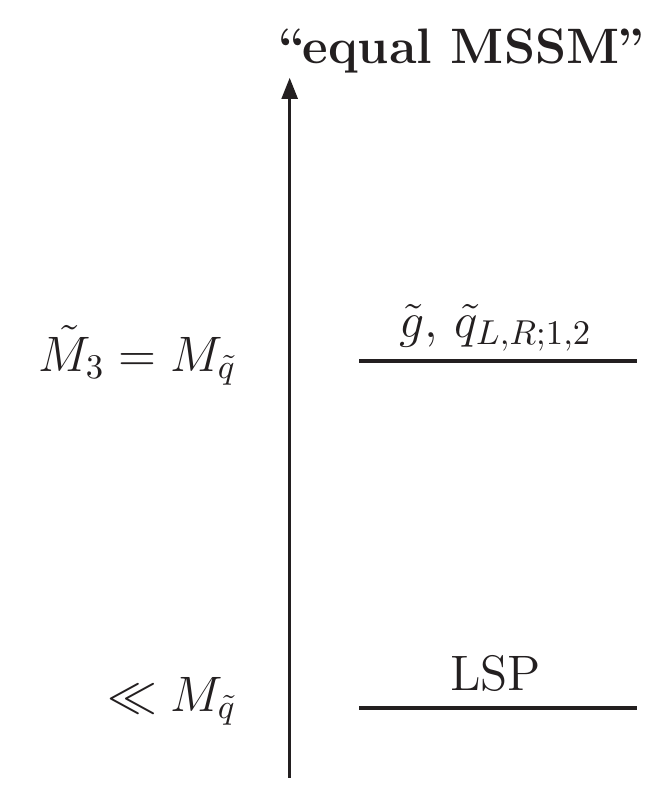}
\includegraphics[width=111pt]{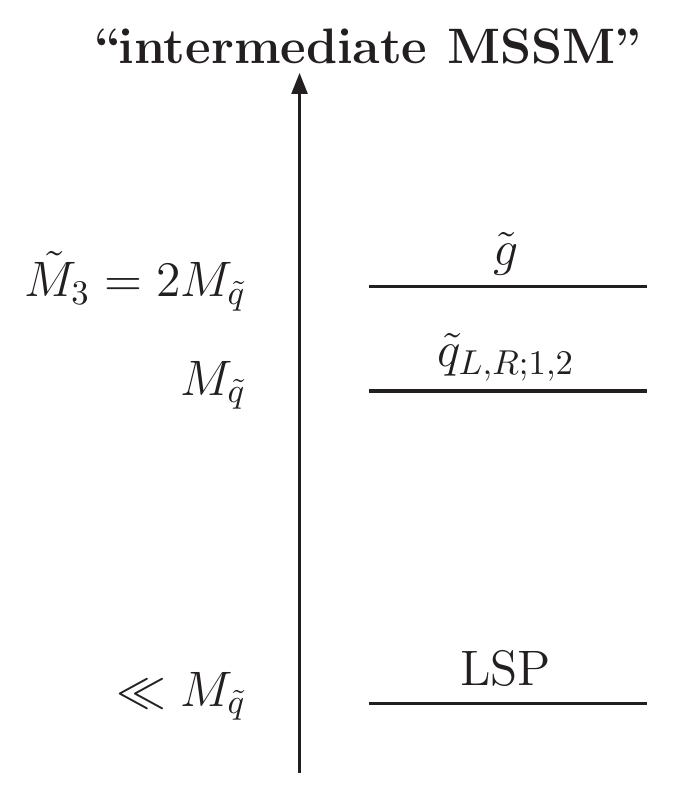}
\includegraphics[width=120pt]{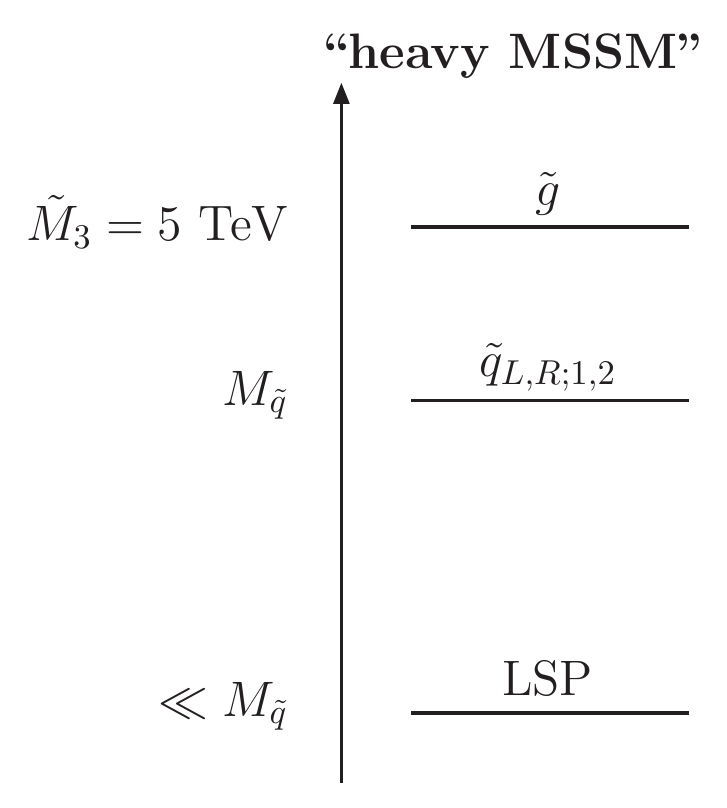}
\caption{The spectra for the simplified models considered in this paper.
The left-most pane illustrates our primary interest -- the
supersoft supersymmetric simplified model (SSSM).  It contains 
a gluino with a large Dirac mass $M_3 = 5$~TeV, 
first and second generation squarks that are roughly $5 \ra 10$ times 
lighter than gluino, and an LSP that is generally assumed to be much
lighter than the squarks.
The three right-most panes illustrate 
the three simplified models of the MSSM to which we compare.
We write the gluino Dirac mass as $M_3$ to be distinguished 
from a Majorana mass written as $\tilde{M}_3$.
Two of the comparison simplified models of the MSSM (``equal MSSM'' and 
``intermediate MSSM'') are designed to provide comparisons 
between typical MSSM spectra and the SSSM\@.  
The third comparison model, ``heavy MSSM'', directly compares
the results for a Dirac gluino versus a Majorana gluino of the
same mass.  Generally the LSP mass is taken to be kinematically
negligible, however we also comment on the relaxation of the bounds 
on the SSSM when the LSP is heavier.}
\label{fig:sssm}
\end{figure*}

\section{Simplified Models and the SSSM}
\label{sec:sssm}

We are interested in calculating the bounds on supersymmetric
models with Dirac gaugino masses.  Our approach is to first 
construct a supersoft supersymmetric simplified model (SSSM)
on which we can apply the $\jetsmissing$ limits from LHC\@.
This is completely analogous to the construction of simplified models
of the MSSM \cite{Alwall:2008ag,Alves:2011wf}, which are now widely
used in presenting the results from LHC searches for supersymmetry.
The SSSM, illustrated in the far left pane of Fig.~\ref{fig:sssm}, 
has a gluino with a large, 
purely Dirac mass, degenerate first and second generation 
squarks (of both handedness), and the lightest supersymmetric particle (LSP)
at the bottom of the spectrum.    
In defining the SSSM, we have explicitly chosen the Dirac gluino mass
to have a fixed large value, $M_3 = 5$~TeV\@.
The large gluino mass implies gluino pair production 
is kinematically forbidden while associated gluino/squark production 
is highly suppressed, leaving squark production as the only potentially
viable colored sparticle production at the LHC\@.
Squarks decay through $\sq \ra q + \mathrm{LSP}$, 
where the quark flavor and chirality depends on the initial squark.  

To perform an apples-for-apples comparison of the constraints 
on supersoft supersymmetry versus the MSSM, we calculate the bounds 
not only on the SSSM, but also three other simplified models of the MSSM\@.
In all of the simplified models, the first and second generation squarks
are degenerate and the LSP is massless.  
The spectra of the three comparison simplified models of the MSSM
are shown in the three right-most panes of Fig.~\ref{fig:sssm}.
The purpose of the comparison models 
is to both validate our analysis against the actual bounds from
experimental analyses (where available), as well as to directly
show the weakness of the bounds on the SSSM in direct contrast 
to the MSSM\@.  
The ``equal MSSM'' and ``intermediate MSSM'' simplified models are 
chosen to provide a comparison with typical MSSM spectra.  
The ``heavy MSSM'' simplified model is highly unnatural within the 
usual MSSM as we have already discussed.  Nevertheless, it illustrates 
the differences in squark mass bounds that remain between a heavy 
Majorana gluino versus a heavy Dirac gluino even when they have  
the same mass. 

Our analyses generally assume the LSP has a kinematically negligible mass.
In the Discussion we also consider the weakening of the bounds as 
the LSP mass is increased.  The LSP could be light gravitino, or 
could instead be some other light neutral superpartner, 
so long as the squark decay proceeds directly to the LSP in 
the one step process $\tilde{q} \ra q + \mathrm{LSP}$.  
We also assume all decays into the LSP are prompt. 
The assumption of short decay chains from heavy squarks 
to a massless LSP implies the bounds we obtain are the 
most optimistic possible using the jets plus missing energy
searches with no leptons in the final state.  

Mapping the bounds from the SSSM onto theories with Dirac gaugino masses 
is straightforward in principle, though model-dependent in practice.
In particular, we do not include electroweak gauginos or Higgsinos
in our spectrum.  
The supersoft supersymmetric model has heavy Dirac gaugino masses, 
with an ordinary MSSM $\mu$-term for the Higgs sector \cite{Fox:2002bu}.  
Several other models incorporate Dirac gauginos 
\cite{Nelson:2002ca,Chacko:2004mi,Carpenter:2005tz,Antoniadis:2005em,Nomura:2005rj,Antoniadis:2006uj,Kribs:2007ac,Amigo:2008rc,Benakli:2008pg,Blechman:2009if,Carpenter:2010as,Kribs:2010md,Abel:2011dc,Frugiuele:2011mh,Itoyama:2011zi}.
In several cases, the gaugino sector approximately preserves
a $U(1)_R$ symmetry, while the Higgs sector does not.
In \cite{Kribs:2007ac} a fully $R$-symmetric supersymmetric model 
was constructed that incorporated not only Dirac gaugino masses but also
$R$-symmetric Higgsino masses.  In this model, additional 
$R$-symmetric contributions to the soft masses were allowed,
and notably, could be nearly arbitrary in flavor-space.
Several phenomenological implications of Dirac gauginos 
as well as fully $R$-symmetric supersymmetry have 
been explored in \cite{Hisano:2006mv,Hsieh:2007wq,Blechman:2008gu,Kribs:2008hq,Choi:2008pi,Plehn:2008ae,Harnik:2008uu,Choi:2008ub,Kribs:2009zy,Belanger:2009wf,Benakli:2009mk,Kumar:2009sf,Chun:2009zx,Benakli:2010gi,Fok:2010vk,DeSimone:2010tf,Choi:2010gc,Choi:2010an,Benakli:2011kz,Kumar:2011np,Heikinheimo:2011fk,Fuks:2012im}.

In this study we do not consider bounds on the third generation squarks.
Third generation squarks receive modifications to their masses
through their interactions with the Higgs supermultiplets.
Given that supersoft supersymmetry has a suppressed $D$-term for 
the Higgs potential, typically this requires heavier stop masses 
as well as separating the scalar masses of the adjoint superfields 
from the corresponding Dirac gaugino masses.  This could be 
accomplished through additional $R$-symmetric $F$-term contributions
to their masses.  In any case, third generation squarks have distinct 
signals involving heavy flavor (with or without leptons), and thus
require incorporating a much larger class of LHC search strategies.
We believe there are interesting differences between the 
third generation phenomenology of a supersoft model versus the MSSM, 
but we leave this for future work.

We also do not consider potentially large flavor-violation in the
squark-gaugino (or squark-gravitino) interactions, as could occur
in an $R$-symmetric model \cite{Kribs:2007ac}.  This would 
add to the heavy flavor component of signals while subtracting
from the $\jetsmissing$ signals that concern us in this paper.
In the interests of demonstrating the differences between the 
SSSM and the simplified models of the MSSM, the latter of which 
cannot have large flavor violation, we do not consider flavor-violation 
in the squark interactions of the SSSM\@.

\section{Aspects of Dirac Gaugino Masses}
\label{sec:advantages}

\subsection{Supersoftness}

A supersoft supersymmetric model contains chiral superfields 
in the adjoint representation of each gauge group of the SM
in addition to the superfields of the MSSM\@.  Supersymmetry
breaking communicated through a $D$-term spurion leads
to Dirac gaugino masses that pair up the fermionic component
from each field strength with the fermionic component of the 
corresponding adjoint superfield.  The adjoint superfields also contain
a complex scalar, whose real and
imaginary component masses are not uniquely determined in 
terms of the Dirac gaugino mass.
The Lagrangian for this setup, in terms of four component spinors, 
is given in Appendix~\ref{app:MRSSMlag}.  

The scalar components of chiral superfields receive one-loop
\emph{finite} contributions to their soft masses from 
gauginos and adjoint scalars, as was shown clearly by \cite{Fox:2002bu}
\begin{eqnarray}
M^2_{\tilde{f}} &=& \sum_i \frac{C_i(f) \alpha_i M_i^2}{\pi} 
                    \log \frac{\tilde{m}^2_i}{M_i^2} \; .
                    \label{eq:scalarmass}
\end{eqnarray}
The sum runs over the three SM gauge groups where $C_i(f)$ 
is quadratic Casimir of the fermion $f$ under the gauge group $i$. 
The $\tilde{m}_i$ are the soft masses for the 
real scalar components of the adjoint superfields.  
The $M_i$ are the Dirac masses for the gauginos.
Assuming the contribution to the squark masses is dominated by 
the Dirac gluino, 
\begin{eqnarray}
\Msq^2 &\simeq& (700~\mathrm{GeV})^2 
                \left( \frac{\Mg}{5~\mathrm{TeV}} \right)^2
                \frac{\log \tilde{r}_3}{\log 1.5}
\end{eqnarray}
where $\tilde{r}_i \equiv \tilde{m}^2_i/M_i^2$.
Somewhat smaller or larger soft masses can be achieved by 
adjusting the ratio $\tilde{r}_3$, since we hold the Dirac 
gluino mass $\Mg = 5$~TeV fixed in the SSSM\@. 

\subsection{Naturalness}

The up-type Higgs mass-squared $m_{H_u}^2$ receives positive 
one-loop finite contributions from the Dirac electroweak gauginos
as well as negative one-loop contributions from the stops.
As was emphasized in Ref.~\cite{Fox:2002bu}, the latter contribution
can easily overwhelm the former, leading to a negative Higgs mass-squared
and thus radiative electroweak symmetry breaking.  Unlike the MSSM,
however, the usual logarithmic divergence from the stop contributions 
to the Higgs mass is cutoff by the Dirac gluino mass, giving
\begin{eqnarray}
\delta m_{H_u}^2 &=& - \frac{3 \lambda_t^2}{8 \pi^2} M_{\tilde{t}}^2 
                       \log \frac{\Mg^2}{M_{\tilde{t}}^2} \; .
\end{eqnarray}
Using Eq.~(\ref{eq:scalarmass}), and approximating 
$\log[\Mg^2/M_{\tilde{t}}^2] \simeq \log[3 \pi/(4 \alpha_s)]$,
we obtain
\begin{eqnarray}
\delta m_{H_u}^2|_{\rm SSSM} &\simeq& 
    - \left( \frac{\Mg}{22} \right)^2 \frac{\log \tilde{r}_3}{\log 1.5} \; .
\end{eqnarray}
Contrast this expression with the analogous one from the 
MSSM \cite{Papucci:2011wy}
\begin{eqnarray}
\delta m_{H_u}^2|_{\rm MSSM} &\simeq& 
    - \left( \frac{\tilde{M}_3}{4} \right)^2 
      \left( \frac{\log \Lambda/\tilde{M}_3}{3} \right)^2 \; .
\end{eqnarray}
where $\tilde{M}_3$ corresponds to the Majorana gluino mass.
This makes it clear that a Dirac gluino can be several times
larger than a Majorana gluino in an MSSM-type model
and yet be \emph{just as natural}, even when comparing against
an MSSM model with a mediation scale that is as low as conceivable, 
$\Lambda \simeq 20 \tilde{M}_3$.  Our choice of Dirac gluino mass 
$\Mg = 5$~TeV with $\tilde{r}_3 \simeq 1.5$ is thus roughly 
equivalent, in the degree of naturalness, 
to a low-scale mediation MSSM model with Majorana
gluino mass $\tilde{M}_3 \simeq 900$~GeV\@.

\subsection{Colored Sparticle Production}

\begin{figure}
\includegraphics[width=0.40\textwidth]{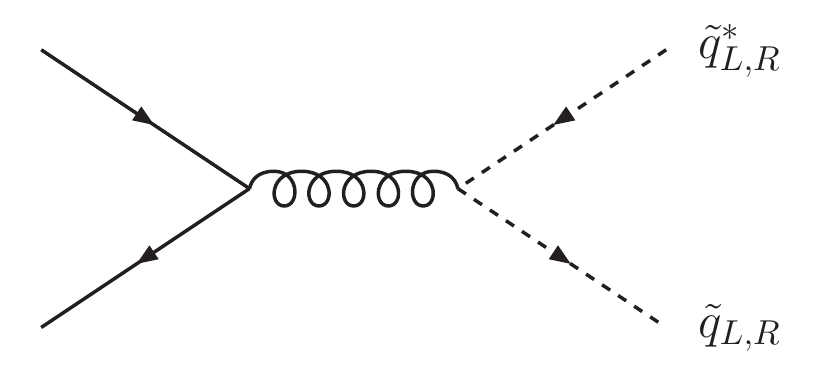}
\includegraphics[width=0.40\textwidth]{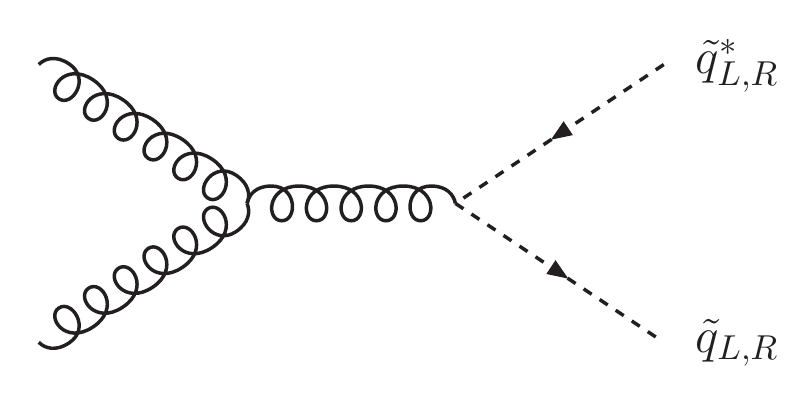}
\includegraphics[width=0.40\textwidth]{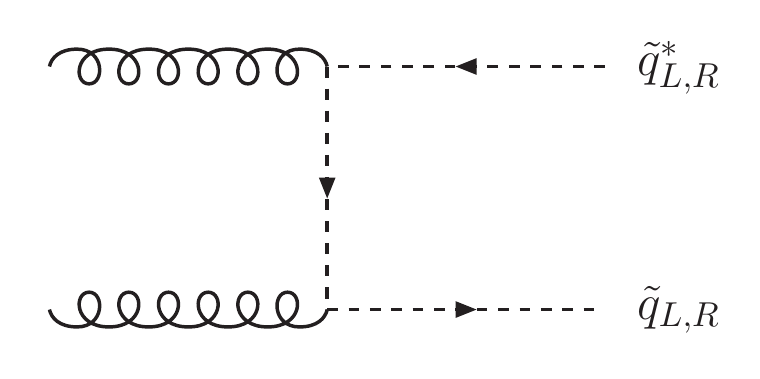}
\caption{The dominant tree level Feynman diagrams for squark production 
at the LHC in the SSSM\@.  Dirac gluino $t$-channel exchange diagrams
(not shown) are suppressed by $1/\Mg^2$ and thus negligible.  
In the MSSM, by contrast, Majorana gluino exchange is suppressed 
by $1/\tilde{M}_3$, and thus relevant even when $\tilde{M}_3$ is large, 
as shown in Fig.~\ref{fig:xsecs}.}
\label{fig:feynman}
\end{figure}

For LHC phenomenology, there are several implications of a 
heavy Dirac gluino.
First, gluino pair production and associated gluino/squark 
production is completely negligible due to the kinematic suppression. 
Squark--anti-squark production can proceed at tree-level through
$gg, q\bar{q} \ra \sq_L\sq^*_L, \sq_R\sq^*_R$,
while the $t$-channel Dirac gluino exchange diagrams are 
suppressed by a factor $1/\Mg^2$.  
There are also mixed-handedness production processes 
$pp \ra \sq_L\sq_R, \sq^*_L \sq^*_R$ through $t$-channel gluino exchange,
but again suppressed by $1/\Mg^2$ in the amplitude.
The contribution of these Dirac gluino exchange diagrams with 
$\Mg = 5$~TeV are at the level of a few percent -- far smaller than
the NLO QCD corrections -- and thus negligible.
The remaining tree-level unsuppressed Feynman diagrams that contribute 
to squark production are shown in Fig.~\ref{fig:feynman}. 
We emphasize that all of these subprocesses require sea quarks
or gluons to initiate at the LHC\@.

The MSSM also contains the same-handedness processes 
$pp \ra \sq_L\sq_L, \sq_R\sq_R$ 
through $t$-channel Majorana gluino exchange, leading to contributions
suppressed by just one power of the gluino mass, $1/\tilde{M}_3$.  
These processes, as well as the mixed-handedness ones 
($pp \ra \sq_L\sq_R, \sq^*_L\sq^*_R$) are initiated by two valence quarks, 
and can lead to a large fraction of the total 
$pp \ra $\,(colored superpartner) cross section.
In the SSSM, the same-handedness processes are simply absent
(no chirality-flipping Majorana mass) while the mixed-handedness 
processes are more suppressed by 
$1/\Mg^2$ instead of $1/\tilde{M}_3$.
This means the cross section for squark production in the SSSM 
can thus be smaller by a factor of $3$ or more
even when comparing the SSSM ($\Mg = 5$~TeV) against the
``heavy MSSM'' simplified model ($\tilde{M}_3 = 5$~TeV).  
Also, the difference between the SSSM and the MSSM grows as 
the squark mass increases, because the final state requires more energy, 
and thus higher partonic $x$, where valence quark distributions dominate 
over gluons or sea quark distributions. 

\subsection{Electroweak inos}

The SSSM, by definition, does not include the effects of the 
Higgsinos or electroweak gauginos.  For general electroweakino masses,
there are two potential effects on our results:
squark cross sections could change due to virtual Higgsino or 
electroweak gaugino exchange; squark decay chains could change 
due to cascades through Higgsinos or electroweak gauginos.  

Higgsino exchange contributions to first and second generation 
squark production is negligible, due to the small Yukawa couplings.  
Electroweak gaugino exchange is suppressed by the smaller
electroweak couplings, and thus not relevant unless the 
electroweak gauginos are significantly lighter than the squarks.
We thus do not expect that our squark production cross section 
calculations to be significantly affected by the Higgsino
and electroweak gaugino spectrum.  

Moreover, while the masses of the electroweak gauginos are 
model-dependent, a supersoft supersymmetric model would predict 
the electroweak gauginos to be $\simeq 4\pi/g$ heavier than sleptons.
Imposing the LEP II bound on slepton masses implies the
electroweak gauginos are generically heavier than the masses
of the squarks we consider in this paper. 
Thus, squark cascade decay through electroweak gauginos is 
kinematically forbidden in supersoft models, and thus we do 
not need to consider it further.

Higgsinos, however, may be lighter than both the squarks and 
the electroweak gauginos.  Naturalness -- obtaining the 
right electroweak symmetry breaking vacuum without 
significant tuning -- certainly favors lighter Higgsinos.
Squark cascade decay through Higgsinos would lead to changes
in the energies of the decay products, as well as the potential addition
of charged leptons or neutrinos in the final state.  
Detailed simulation of these cases is highly model-dependent.  
Nevertheless, the jets plus missing energy bounds on models with 
lighter Higgsinos could be substantially weaker 
if the average hadronic activity is reduced.  
On the other hand, the bounds from other supersymmetric searches 
could be substantially stronger if the squark cascade 
through Higgsinos results in hard leptons or photons.
We note however that searches more specific to models with 
Majorana neutralinos, such as same-sign lepton final states, may not 
yield strong bounds if the model is approximately $R$-symmetric,
and so again we are left to model-dependent investigations
to make quantitative statements.

\section{Recasting LHC Limits}
\label{sec:lhclimits}

To recast LHC limits on colored superparticle production into the SSSM, 
we follow the analyses searching for supersymmetry through
$\jetsmissing$ signals performed by 
ATLAS \cite{atlas_jetsmet} and 
CMS \cite{cms_alpha_t,cms_mht,razor}.
Of the existing supersymmetry searches, jets plus missing energy is 
the simplest, and involves the fewest assumptions about the spectrum.

To simulate the supersymmetric signal, we use 
PYTHIA6.4 \cite{Sjostrand:2006za}; the first and 
second generation squarks are set to have equal mass, the gravitino is 
chosen to be the LSP, and all other superpartners are 
decoupled (set to $5$~TeV). We use CTEQ6L1 parton distribution 
functions, generating a sufficient number of events such that 
statistical fluctuations have negligible effect on our results. To 
incorporate detector effects into our signal simulations, all events are 
passed through the Delphes~\cite{Ovyn:2009tx} program using ATLAS or CMS 
detector options and adopting the corresponding experiment's 
jet definitions:  
anti-$k_T, R = 0.4$ for the ATLAS search \cite{atlas_jetsmet}, and 
anti-$k_T, R = 0.5$ for the CMS searches \cite{cms_alpha_t,cms_mht,razor}. 
We repeat the same steps for the 
three simplified models of the MSSM
(c.f.~Fig.~\ref{fig:sssm}) allowing all 
combinations of $\sq\sq$, $\sq^* \sq^*$, $\sq \sq^*$ 
as well as gluino pair production and associated squark 
plus gluino production.  Note that our ``heavy MSSM'' simplified model 
is an existing CMS simplified model, 
``T2'' \cite{alphat_supplementalplots}.

\begin{figure}
\centering
\includegraphics[width=0.48\textwidth]{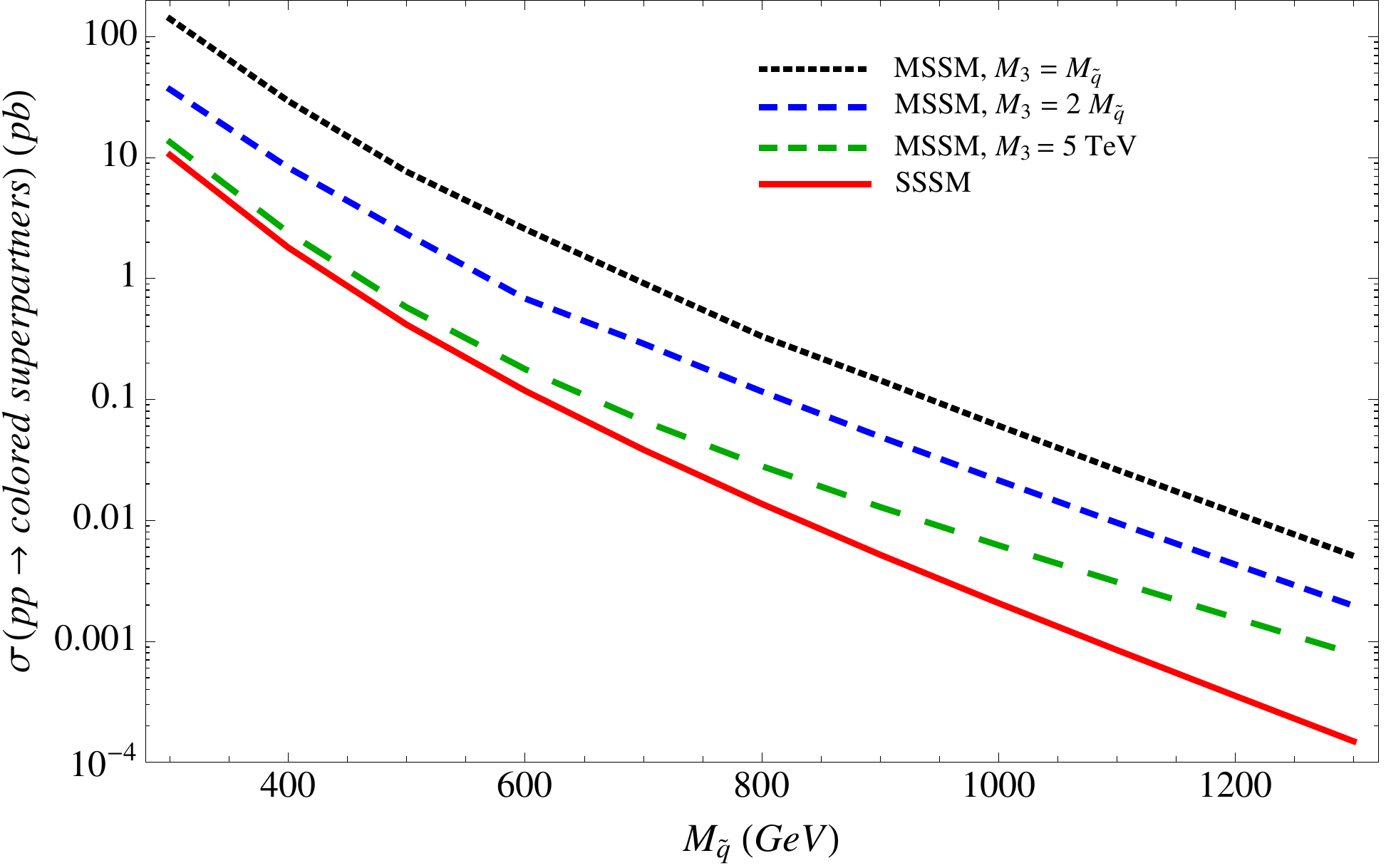}
\caption{Cross sections at the 7 TeV LHC for colored superpartner 
production. The four lines correspond to the four simplified models
shown in Fig.~\ref{fig:sssm}, where the first and second generation 
squarks are degenerate with mass $\Msq$.  The solid line 
shows the cross section for the SSSM where the cross section is
dominated by $\sq\sq^*$ final states, while the dashed lines show 
cross sections for the three simplified models of the MSSM\@. 
All cross sections are calculated to 
next-to-leading order using PROSPINOv2.1~\cite{prospino}, CTEQ6L1 parton 
distribution functions,  and default scale choices. For event generation, 
we use PYTHIA6.4~\cite{Sjostrand:2006za} and rescale the cross section to 
match those shown here.}
\label{fig:xsecs}
\end{figure}

Colored superpartner production cross sections receive sizable 
next-to-leading order (NLO) corrections. 
To incorporate these corrections, we feed the spectra into 
PROSPINO~\cite{prospino}, restricting the processes appropriately
for each simplified model (i.e., just $pp \ra \sq \sq^*$ for the SSSM).  
The cross sections are shown in Fig.~\ref{fig:xsecs} for each of
the simplified models as a function of squark mass.  
Depending on the scale choice and the squark mass, 
we find the $K$-factor ranges from $1.7$-$2.1$.  
This takes into account the increased rate at NLO,
through not the kinematic distribution of events.  

The analyses we are interested 
in~\cite{atlas_jetsmet,cms_alpha_t,cms_mht,razor}, 
are broken up into several channels. For some analyses the channels 
are orthogonal, while in other analyses one event can fall into multiple 
channels. To set limits we begin by counting the number of supersymmetry 
events in each analysis channel for several squark masses. The number of 
supersymmetric events passing cuts is translated into a 
mass-dependent acceptance for each channel. We then form the $95\%$ CL 
limit, using the likelihood ratio test statistic~\cite{conway1}:
\begin{align}
0.05 &= \frac{\int_0^{\infty} db' \sum_0^{N_{i,obs}}  \frac{(\mu_{i,b} + \mu_{i,s})^{N_{i, obs}} e^{-(\mu_{i,b} + \mu_{i,s})} }{(N_{i, obs})!} G(\mu_b, b') }{\int_0^{\infty} db' \sum_0^{N_{i,obs}} \frac{\mu_b^{N_{i, obs}} e^{-\mu_b} }{(N_{i, obs})!} G(\mu_b, b') }.
\label{eq:lr}
\end{align}
Here $\mu_{i,b} \equiv N_{i, exp}$ is the number of expected 
SM background events and $\mu_{i,s} \equiv N_{i, SUSY}$ is the 
number of signal events. To estimate the effects of systematic errors, 
the number of SM events is modulated by a Gaussian weighting 
factor~\cite{conway}. 
Specifically, we shift $\mu_b \rightarrow \mu_b(1+f_b)$, where $f_b$ 
is drawn from a Gaussian distribution centered at zero and with 
standard deviation $\sigma_f = \sigma_{i,SM}/N_{i, exp}$, where 
$\sigma_{i,SM}$ is the quoted systematic uncertainty 
(taken directly from~\cite{cms_alpha_t,cms_mht,razor,atlas_jetsmet}). 
Whenever the systematic error is asymmetric, we use the larger 
(in absolute value) number. To combine channels (when appropriate), 
we simply replace the right-hand side of Eq.~(\ref{eq:lr}) with the product 
over all channels.

The number of supersymmetry events in a 
particular channel is the product of the cross section, luminosity, 
acceptance and efficiency,
\begin{equation}
N_{i, SUSY} = \mathcal{L} \cdot K(\Msq) \, \sigma(\Msq) \cdot 
                A(\Msq) \cdot \epsilon, 
\label{eq:nsusy}
\end{equation}
where $K(\Msq)$ is the mass-dependent $K$-factor to account
for the larger rate at NLO\@. Within our simplified setup, the only 
parameter the cross section and acceptance depend upon is the mass of the 
squark -- thus Eq.~(\ref{eq:lr}) is simply a limit on the squark mass. 

While the likelihood ratio test statistic is particularly well suited 
to analyses with low event counts, it is just one possibility. 
To test that our results do not depend on this choice, we have also 
computed limits using the sum-$\chi^2$ test statistic,
\begin{equation}
\sum_{i=1}^{chan} \frac{(N_{i,obs} - (N_{i,exp} + N_{i,SUSY}) )^2}{N_{i,exp} + N_{i,SUSY} + \sigma^2_{i,SM}}.
\label{eq:chisq}
\end{equation}
We find our results using different test statistics are broadly consistent, 
with the biggest differences being, as expected, when the number 
of events in a particular channel is low.

The constructions in Eq.~(\ref{eq:lr}) and (\ref{eq:chisq}) are only 
approximate. Both formulations assume a Gaussian treatment of the 
systematics is appropriate, and correlations among uncertainties when 
combining channels are completely neglected.  A more complete
treatment of the correlated experimental uncertainties may be
possible through RECAST \cite{Cranmer:2010hk}, which we leave
for future work.  

\begin{table*}[t]
\centering
\begin{tabular}{|c|c|c|c|c|}\hline
               &  SSSM            & ``equal MSSM''         & ``intermediate MSSM''            & ``heavy MSSM'' \\ 
search channel &  ($M_3 = 5$~TeV) & ($\tilde{M}_3 = \Msq$) & ($\tilde{M}_3 = 2 \times \Msq$) & ($\tilde{M}_3 = 5\,\tev$) \\ \hline\hline
ATLAS jets + $\slashed E_T$  & & & &  \\
4.7 $\fb^{-1}$ & & & &  \\ \hline
SRA (2j) medium  & 737 GeV & 1245 GeV & 1096 GeV & 890 GeV \\
SRA (2j) tight & 634 GeV & {\bf 1453 GeV} & {\bf 1305 GeV} & {\bf 1063 GeV} \\
SRA$'$\,(2j) tight & {\bf 748 GeV} & 1189 GeV & 1061 GeV & 861 GeV \\
SRB (3j) tight & 537 GeV & 1342 GeV & 1202 GeV & 848 GeV \\
SRC (4j) loose & 566 GeV & 973 GeV & 770 GeV & 584 GeV \\
SRC (4j) medium & 634 GeV & 1095 GeV & 894 GeV & 670 GeV \\
SRC (4j) tight & ** & 1082 GeV & 831 GeV & 431 GeV \\
SRD (5j) tight & 383 GeV & 1076 GeV & 803 GeV & 484 GeV \\
SRE (6j) loose & ** & 731 GeV & 500 GeV & 328 GeV \\
SRE (6j) medium & 491 GeV & 979 GeV & 712 GeV & 521 GeV \\
SRE (6j) tight & ** & 933 GeV & 634 GeV & 388 GeV \\ \hline \hline
CMS $\alpha_T$ & & & & \\ 
1.14 $\fb^{-1}$  & & & & \\ \hline
$H_T \in [275, 325)\,\gev$ & 396 GeV  & 528 GeV & 489 GeV & 399 GeV   \\ 
$H_T \in [325, 375)\,\gev$ & 454 GeV  & 594 GeV & 533 GeV & 473 GeV \\
$H_T \in [375, 475)\,\gev$ & 509 GeV & 698 GeV & 631 GeV & 548 GeV \\ 
$H_T \in [475, 575)\,\gev$ & {\bf 540 GeV} & 786 GeV & 694 GeV & {\bf 570 GeV} \\ 
$H_T \in [575, 675)\,\gev$ & 487 GeV & 859 GeV & 770 GeV & 565 GeV  \\
$H_T \in [675, 775)\,\gev$ & 373 GeV & 932 GeV & 833 GeV & 460 GeV \\
$H_T \in [775, 875)\,\gev$ & ** & 960 GeV &  806 GeV & ** \\
$H_T \ge 875\,\gev$ & ** & {\bf 1160 GeV} & {\bf 968 GeV} & ** \\ \hline
{\bf combined} & 684 GeV & 1178 GeV & 1032 GeV  & 786 GeV  \\ \hline \hline
CMS jets + MHT  & & & & \\
 1.1 $\fb^{-1}$ &  & & & \\ \hline
$\slashed E_T > 350\,\gev, H_T > 500\,\gev$ & {\bf 593 GeV} & 989 GeV & 844 GeV & 648 GeV \\
$H_T > 500\,\gev$ & 500 GeV  & 989 GeV & 799 GeV  & 563 GeV \\
$\slashed E_T > 500\,\gev, H_T > 800\,\gev$ & 416 GeV  & {\bf 1154 GeV} & {\bf 981 
GeV}  & {\bf 661 GeV}  \\  \hline \hline
CMS \emph{razor}, & & & & \\ 
4.4 $\fb^{-1}$ & &  & & \\ \hline
0 $\ell$, S1 & ** & 639 GeV  & ** & **  \\ 
0 $\ell$, S2 & ** & ** & ** & **  \\ 
0 $\ell$, S3 & **  & 960 GeV & 783 GeV & 434 GeV\\ 
0 $\ell$, S4 & ** & {\bf 1082 GeV} & {\bf 898 GeV} & 349 GeV \\ 
0 $\ell$, S5 & 485 GeV & 779 GeV & 653 GeV & 514 GeV \\
0 $\ell$, S6 & {\bf 505 GeV} & 794 GeV & 690 GeV & {\bf 556 GeV}\\ \hline
{\bf combined} & 588 GeV & 1137 GeV & 961 GeV & 677 GeV \\\hline 
\end{tabular}
\caption{Channel-by-channel and combined observed limits on the simplified 
models illustrated in Fig.~\ref{fig:sssm} using the likelihood ratio 
test statistic. 
Channels marked with an asterisk have limits lower than $300$~GeV, 
while the strongest channel for a given analysis is shown in bold. 
Combined limits are shown for analyses where the individual channels 
are orthogonal.}
\label{tab:bigtab_poisson}
\end{table*}

The exact limits we can place from the experimental analyses depend 
on several factors.  The luminosity and the systematic uncertainties 
on the background are examples of factors that evolve with time, 
while the signal cross section and acceptance (for a given analysis) 
are fixed.  To make our study as general as possible, we show our
derived acceptances as a function of squark mass in a series
of Figures in Appendix~\ref{app:acceptance}.  
These numbers allow us to estimate limits as the
luminosity increases, at least for a fixed analysis strategy.
Nevertheless, we do calculate the squark mass limits using the 
experiments' quoted luminosities and background uncertainties 
in Table~\ref{tab:bigtab_poisson}.  Both the limits from individual 
channels, as well as combined limits (in cases where the channels are
distinct and nonoverlapping) are given. The cross sections have already 
been shown in Fig.~\ref{fig:xsecs}, leaving the derived acceptance 
times efficiency as the only undetermined factor in 
Eq.~(\ref{eq:nsusy}). 

In the following subsections we present the set of analyses used to bound 
the parameter space of our SSSM\@. 
The details of the analyses cuts can be found in 
Refs.~\cite{atlas_jetsmet,cms_alpha_t,cms_mht,razor}.  
For ease of comparison, all of the bounds we obtain for each 
analysis strategy from each experiment are presented in 
Table~\ref{tab:bigtab_poisson}.  The table provides the bounds
on the SSSM, as well as the three simplified models of the MSSM 
shown in Fig.~\ref{fig:sssm}. 
In the following, we discuss the important observables for each analysis, 
then describe our extracted limits.

\subsection{ATLAS Limits with 4.7~fb$^{-1}$}
\label{sec:ATLlim}

The first analysis we consider is the ATLAS jets plus missing energy search 
performed in Ref.~\cite{atlas_jetsmet}. Events with no leptons and 
large missing energy 
are subjected to several subanalyses, each with a different jet 
multiplicity requirement (2-6 jets).
Within each multiplicity subanalysis, cuts are then placed on the 
individual jet transverse momenta, the effective 
mass for a given multiplicity: 
$m_{eff}(N) = \sum_{i=1}^{N} p_{T,i} + \slashchar{E}_T$, and the 
ratio of missing energy to effective mass. To further reduce backgrounds 
from poorly measured QCD jets, a cut is also placed on the minimum 
azimuthal angle between the missing momenta vector and any 
(sufficiently hard) jet. Surviving events are then classified 
according to their inclusive $m_{eff}(inc.)$, which differs from 
$m_{eff}(N)$ in that all jets with $p_T > 40$~GeV are included 
in the sum. The $m_{eff}(inc.)$ classifications are referred to as 
``loose'', ``medium'' and ``tight''. 
There are eleven total channels, since not every jet multiplicity 
has all three $m_{eff}(inc.)$ classifications.

The derived $A(\Msq)\cdot \epsilon$ for the $11$ different 
channels in the ATLAS jets plus missing energy search~\cite{atlas_jetsmet} 
are shown in Fig.~\ref{fig:atlas_limits} in Appendix~\ref{app:acceptance}. 
We show the acceptance times efficiency as a function of squark mass 
both in the SSSM as well as the simplified models of the MSSM\@.

We emphasize that Fig.~\ref{fig:atlas_limits} only gives a piece of the 
limit calculation -- a large efficiency does not necessarily mean a good 
limit, as the background may also be large. Applying Eq.~(\ref{eq:chisq}) ¶
using the observed event counts from Ref.~\cite{atlas_jetsmet}, 
we find the $2$-jet (A, A$'$) channels have the best sensitivity: 
$\Msq > 737$~GeV and $\Msq > 748$~GeV respectively 
(95\% CL, see Table~\ref{tab:bigtab_poisson} for full details). 
For the simplified models of the MSSM, we find the bounds range
from $\Msq > 1063$~GeV (for the ``heavy MSSM'' simplified model)
to $\Msq > 1453$~GeV (for the ``equal MSSM'' simplified model). 
The acceptance/efficiency factors for the different scenarios are similar,
as shown in Fig.~\ref{fig:atlas_limits} in Appendix~\ref{app:acceptance},
and thus the difference in the limits is driven by the 
larger cross sections in the simplified models of the MSSM\@. 

There is a another ATLAS supersymmetry search focusing on \emph{very} high 
jet multiplicity, $\ge 6$ jets~\cite{Aad:2011qa}. This search 
is most sensitive to supersymmetric events with long decay chains, such as 
from gluino pair production. For events dominated by short decay chains, 
i.e., the SSSM, we expect the high-multiplicity tails are not large enough 
to be seen over the background uncertainty.  
We verified this by passing the SSSM through the analysis strategy 
following Ref.~\cite{Aad:2011qa}, where we find the limits are indeed poor 
in comparison to the other strategies, and so we do not present them.

\subsection{CMS Limits with $\sim$~1-5~fb$^{-1}$}
\label{sec:cmslim}

We now turn to supersymmetry searches performed by the CMS collaboration. 
We follow three different jets plus $\slashchar{E}_T$ search strategies. 
The first, in Ref.~\cite{cms_alpha_t}, uses the $\alpha_T$ variable to 
distinguish signal -- events with real missing energy -- from background 
events where the missing energy comes from mismeasurement. The second, 
Ref~\cite{cms_mht} relies on large $\slashchar{E}_T$ and $H_T$ to suppress 
background, while the third uses the so-called \emph{razor} variables 
developed in~\cite{Rogan:2010kb}. We follow the same procedure as in 
Sec.~\ref{sec:ATLlim}; we derive $A(\Msq)\cdot \epsilon$ 
for each analysis using Monte-Carlo events, then follow Eq.~(\ref{eq:lr}) 
to set limits on the squark masses. 
The $A(\Msq)\cdot \epsilon$ curves depend only on the 
analysis cuts and can be applied unchanged to future data sets 
with increased luminosity or improved background modeling.

\subsubsection{Search based on $\alpha_T$, $1.1\,\fb^{-1}$}
\label{ss:alphat}

In addition to basic identification cuts, this analysis requires that the 
leading two jets have $p_T > 100$~GeV and that the leading jet lies 
within the tracker. After vetoing events with leptons or photons, events 
are binned according to their overall $H_T = \sum_{i}^{jets} E_{T,i}$, 
starting with $H_T = 275$~GeV: 
two $50$~GeV bins spanning up to $375$~GeV, 
four $100$~GeV bins, 
then one bin containing all events with $H_T > 875$~GeV\@.

The hadronic activity in each event is massaged into two 
pseudojets\footnote{For events with only two jets this massaging is 
trivial. For events with multiple jets, the jets are combined until the 
event contains only two 'pseudojets;. The choice of how the jets are added 
is determined by minimizing the difference between the scalar sum of the 
jet $E_T$ between the two pseudojets. See Ref.~\cite{cms_alpha_t}.} which 
are used to calculate $\alpha_T$, defined as
\begin{equation}
\alpha_T = \frac{E_{T,2}}{M_{T, jj}}.
\end{equation} 
Cutting at $\alpha_T > 0.55$, the pure QCD contribution to the background 
becomes highly suppressed.

The acceptance times efficiency derived for each channel of this analysis 
is shown in Fig.~\ref{fig:cms_limits} in Appendix~\ref{app:acceptance}.
The squark mass directly sets the net transverse energy in an event, 
so the peak efficiency in a particular $H_T$ bin simply tracks the 
squark mass.

As before, the channel-by-channel limits from this search are shown in 
Table~\ref{tab:bigtab_poisson}. 
For the SSSM and the ``heavy MSSM'' simplified model, 
the most sensitive channels are for mid-range $H_T$ where the signal 
rate is still large and the background uncertainties are falling. 
In comparison, 
the cross section for the other two simplified models of the MSSM 
falls much slower with increasing squark mass, propped up by the 
lighter Majorana gluinos, leading to the highest $H_T$ bins being 
the most constraining.  As the different $H_T$ channels are orthogonal, 
it is straightforward to combine them, resulting in a better limit.  
Forming the product of likelihood ratios over all channels, 
we find an observed $95\%$\,CL limit of 
$\Msq \gtrsim 684$~GeV for the SSSM, and 
$\Msq \gtrsim 786$~GeV for the ``heavy MSSM'' simplified model.
The latter limit is in good agreement with the observed limit shown 
for the ``T2'' simplified model in Ref.~\cite{alphat_supplementalplots}, 
giving confidence that we have successfully reproduced their analysis.
For the other cases, the combined limits are much higher: 
$\Msq \gtrsim 1160$~GeV for the ``equal MSSM'' simplified model, and 
$\Msq > 1032$~GeV for the ``intermediate MSSM'' simplified model.

\subsubsection{Search based on $\slashchar{E}_T, H_T$, $1.1\,\fb^{-1}$}
\label{ss:cmsjetsmissing}

The second CMS search strategy we consider is more traditional 
in that it is based simply on large multiplicity of high-$p_T$ jets 
and large missing energy (see Ref.~\cite{cms_mht}). 
At least three jets of $p_T > 50$~GeV, $|\eta| < 2.5$ 
are required and no leptons ($p_T > 10$~GeV, $|\eta| < 2.5$) are permitted. 
Selected events require a minimum $\slashchar{E}_T$ or $200$~GeV 
sufficiently separated from the jets\footnote{All jets $p_T > 30$~GeV, 
$|\eta| < 5.0$ are used in the $\slashchar{E}_T$ calculation, and the 
minimum separation is $\Delta \phi(j_i, \slashchar{E}_T)\,> 0.5$ for the 
hardest two jets and $0.3$ for the third hardest jet.} and a minimum 
$H_T = 350$~GeV\@. Passing events are piled into three 
further categories depending on $\slashchar{E}_T$, $H_T$: 
i.)   $\slashchar{E}_T > 350$~GeV, $H_T > 500$~GeV, 
ii.)  $H_T > 800$~GeV, 
iii.) $\slashchar{E}_T > 500$~GeV, $H_T > 800$~GeV\@.

The acceptance/efficiency factors we find for the three channels are shown 
in Fig.~\ref{fig:cms_MHT} in Appendix~\ref{app:acceptance}.
This analysis gives has a similar trends to the previous search. The 
channel with lowest $\slashchar{E}_T$ and $H_T$ are most constraining for 
the supersoft and heavy-gluino MSSM, while the channel with the tightest 
cuts are more stringent for the light gluino MSSM scenarios. The strongest 
individual channel limits are quite similar to the 
$\alpha_T$ case. As the channels are not orthogonal we do not combine 
and simply quote the strongest individual channel: 
$\Msq >  593$~GeV (SSSM);
$\Msq > 1154$~GeV (``light MSSM'' simplified model);
$\Msq >  981$~GeV (``intermediate MSSM'' simplified model);
$\Msq >  661$~GeV (``heavy MSSM'' simplified model).

\subsubsection{Search based on \emph{razor} variables}
\label{ss:razor}

The final CMS search strategy we consider from Ref.~\cite{razor}
utilizes the \emph{razor} variables to discriminate signal 
from background. For the \emph{razor} analysis, all objects passing basic 
identification and selection cuts are grouped into two ``mega-jets''. 
The division of particles into mega-jets is determined by which combination 
yields mega-jets that are closest in invariant mass. Once the mega-jets 
are formed, one boosts longitudinally to the frame where the two 
mega-jets have equal and opposite momenta along the beam direction ($p_z$). 
In this special frame, one calculates $M^R_T$ and $M_R$ defined as:
\begin{eqnarray}
(M^R_T)^2 &=& \frac{1}{2} \Big( \slashchar{E}_T\, (p_{T,j_1} + p_{T,j_2}) - 
\vec{\slashchar{E}}_T\cdot(\vec{p}_{T,j_1} + \vec{p}_{T,j_2} ) \Big), 
\nonumber \\
M_R &=& \sqrt{(E_{j_1} + E_{j_2})^2 - (p_{z, j_1} + p_{z,j_2})^2}.
\end{eqnarray}
The magnitude of $M_R$ and the ratio $R^2 = (M^R_T/M_R)^2$ are then used 
to differentiate signal and background. The cut values for $R^2$ and $M_R$ 
depend on whether the event contains any isolated electrons or muons.  For 
our signal, events with isolated leptons are rare, so we focus on the 
hadronic channel. The events in each channel are divided up into several
bins then compared to the background, which has been extrapolated from 
a signal-free ``fit-region''. 

To set limits, we considered the six analysis regions defined by
Ref.~\cite{razor}: 
\begin{eqnarray}
S_1 & :\, & R^2 \in [0.18, 0.3], M_R \in [2.0\,\tev, 3.5\,\tev] \nonumber \\
S_2 & :\, & R^2 \in [ 0.3, 0.5], M_R \in [2.0\,\tev, 3.5\,\tev] \nonumber \\
S_3 & :\, & R^2 \in [0.18, 0.5], M_R \in [1.0\,\tev, 2.0\,\tev] \nonumber \\
S_4 & :\, & R^2 \in [ 0.3, 0.5], M_R \in [1.0\,\tev, 2.0\,\tev] \nonumber \\
S_5 & :\, & R^2 \in [ 0.2, 0.3], M_R \in [650\,\gev, 1.0\,\tev] \nonumber \\
S_6 & :\, & R^2 \in [ 0.4, 0.5], M_R \in [400\,\gev, 1.0\,\tev] \nonumber 
\end{eqnarray}
The acceptance times efficiency factor for each channel as a function of 
squark mass is shown in Fig.~\ref{fig:cms_razor} in 
Appendix~\ref{app:acceptance}.

Since the six regions are orthogonal, we can combine channels, 
leading to the limits: 
$\Msq > 588$~GeV (SSSM), 
$\Msq > 677$~GeV (``heavy MSSM'' simplified model), and 
$\Msq \gtrsim 1$~TeV for the ``equal MSSM'' and ``intermediate MSSM''
simplified models with lighter Majorana gluinos. 
For the simplified models with heavy gluinos, the colored superpartner 
cross section falls fastest with increasing $\Msq$, 
and thus the bounds are dominated by the the lowest $M_R$ bins. 
As the gluino mass decreases, the superpartner cross section 
falls less precipitously, and the larger $M_R$ bins provide stronger 
constraints. 

The limits we set with the 6-bin approach are conservative estimates. 
Utilizing an unbinned likelihood approach (as done in Ref.~\cite{razor}),
our limits may improve. However, the unbinned approach requires a complete, 
smooth description of the background (and signal) in the two-dimensional 
($R, M_R$) plane and makes our limit more sensitive to details of the 
detector modeling and correlations among systematics. 

\section{Luminosity Extrapolation}
\label{sec:lum}

It is interesting to extrapolate the squark mass limits set in the previous 
section out to higher luminosity. Since we do not have the observed data 
from the future, we extrapolate using the expected limit, meaning $N_{i, obs}$ 
is set equal to $N_{i, exp}$ in Eq.~(\ref{eq:lr}). As we want to vary the 
luminosity, the background number of events is actually $N_{i, ex} \times 
(\mathcal L/\mathcal L_0)$ where $\mathcal L_0$ is the luminosity used to 
derive efficiencies (the luminosity 
in~\cite{atlas_jetsmet,cms_alpha_t,cms_mht,razor}), 
and $\mathcal L$ is the projection luminosity. This 
extrapolation is conservative in that it assumes there is no 
re-optimization of the analysis cuts and that the systematic uncertainties 
remain unchanged.

\begin{figure}[t]
\centering
\includegraphics[width=3.5in]{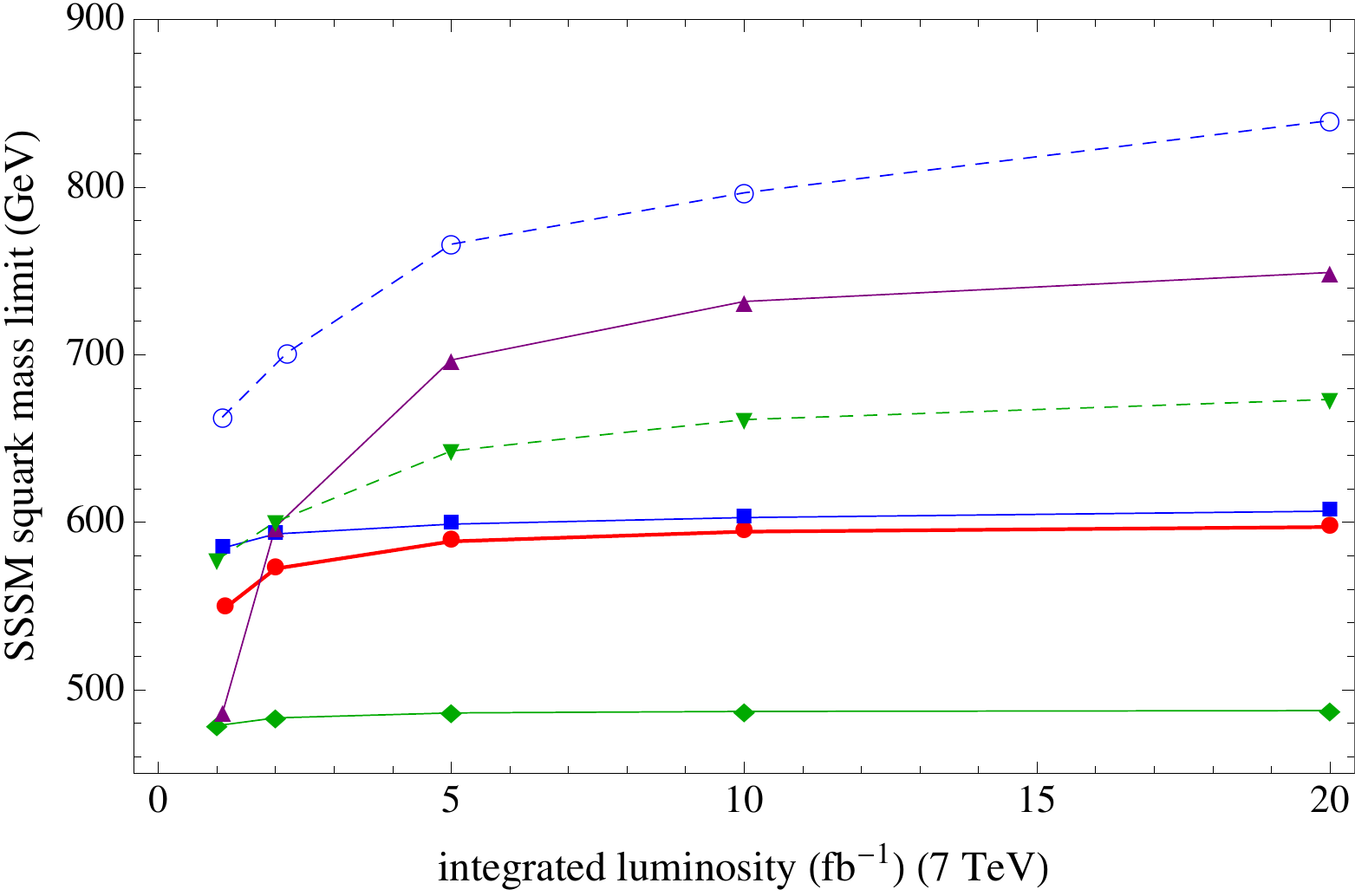}
\caption{Projection of the expected limits to larger integrated 
luminosity holding the analysis strategy fixed as well as 
$\sqrt{s} = 7$~TeV\@.  For each detector analysis strategy, 
the strongest individual channel  
is shown, while for the $\alpha_T$ and \emph{razor} analyses we show the 
projection of the combined channel limit as well.
The red line corresponds to CMS jets plus $\slashchar{E}_T$, 
the blue corresponds to CMS $\alpha_T$ (solid is the single channel limit, 
dashed is the combined limit), 
green (solid and dashed) shows CMS \emph{razor}, and 
purple is ATLAS jets + $\slashchar{E}_T$.
We emphasize that we have plotted only the \emph{expected} limits, 
to be distinguished from the \emph{observed} limits that we show 
in Table~\ref{tab:bigtab_poisson}.  The small differences between
the expected and observed limits are at roughly the 10\% level, 
characteristic of background fluctuations.}
\label{fig:projsingle}
\end{figure}

We perform an extrapolation using the individual channel with the 
strongest limits from the various analyses, as well as the combined
channels for the CMS $\alpha_T$ strategy and the CMS \emph{razor}
strategy. These extrapolations are shown in Fig.~\ref{fig:projsingle}.
As the luminosity increases, we find the limits on the squark mass 
do not improve dramatically.
The CMS $\alpha_T$ search appears to be the best performing 
future search on the SSSM, with improvements on the squark mass 
bounds of expected to be roughly $15$-$25\%$.
The limits asymptote fairly quickly once the analyses become dominated 
by systematic uncertainties rather than by statistical uncertainties. 
If the background systematics improve in the future, these projections 
could easily be redone using the signal acceptance times 
efficiency curves shown earlier.

\section{Discussion}
\label{sec:discussion}

We have shown that our simplified model of supersoft supersymmetry 
is clearly much less constrained by LHC searches for supersymmetry 
than comparable simplified models of the MSSM\@.  
We find the bounds on first and second generation
squark masses in the SSSM to be between $680$ to $750$~GeV, 
depending on the experiment, the particular search strategy,
and the amount of integrated luminosity analyzed. 
This is fully consistent with the one-loop finite mass generated
from a $5$~TeV Dirac gluino (with $\tilde{r}_3 \simeq 1.5$),
as we showed in detail in Sec.~\ref{sec:advantages}.  
Importantly, these bounds are only modestly improved with the 
increased luminosity of the LHC\@.  
We emphasize that our luminosity extrapolation was done assuming
the search strategies were \emph{unchanged}, and applied to more 
luminosity at $\sqrt{s} = 7$~TeV\@.  Nevertheless, the clear
conclusion from the extrapolation is that the SSSM with a 
kinematically inaccessible Dirac gluino production remains
safe from LHC bounds now and into the near future.

One of the more striking results is that the CMS $\alpha_T$ analysis 
provided the strongest bound on the squark masses of the SSSM
at $1$~fb$^{-1}$.  The ATLAS jets plus missing energy search strategy, 
despite the considerable integrated luminosity $4.7$~fb$^{-1}$,  
resulted in only a slightly better bound.  Our interpretation of these
results is the $\alpha_T$ search, which was designed to maximize
signal over background with $2$ jets plus missing energy,
provides an ideal search strategy for the SSSM\@.
This is due in large part because the $\alpha_T$ strategy 
implements a wide range of search channels at intermediate values 
of $H_T$ that are precisely within the range expected 
for $\sim 600 \ra 800$~GeV squarks of the SSSM\@.  
This is also borne out by the best bound from the CMS MHT strategy
being the lower missing energy, lower $H_T$ channel (distinctly different
from the simplified models of the MSSM with lighter gluinos).
Examining the expected limits from Fig.~\ref{fig:projsingle},
we see that the $1$~fb$^{-1}$ CMS $\alpha_T$ strategy is expected
to yield the same bound on squarks in the SSSM as about a 
$4$~fb$^{-1}$ jets plus missing energy ATLAS analysis.
This appears to be because the 2 jet search strategies done by ATLAS 
require very large $m_{eff}$.  Indeed, the ATLAS channel with the 
best bound on the SSSM (SRA$'$) had the \emph{least restrictive} 
cut on $m_{eff}$ (greater than $1200$~GeV).  
Similarly, the CMS \emph{razor} analysis appears to be best optimized for 
very high mass superpartner searches.

Our study focused on a nearly massless LSP, so that we could
perform an apples-for-apples comparison between the LHC bounds 
on simplified models of the MSSM versus the SSSM\@.
It is interesting to consider how the bounds change as 
the LSP mass is increased.  Since the strongest expected bound on 
the squark masses in the SSSM comes from the CMS $\alpha_T$ analysis, 
we explored a variation of the SSSM where we allowed the 
LSP mass within the range $0 \le m_{\rm LSP} \le 300$~GeV\@.
We find CMS $\alpha_T$ limits for $m_{\rm LSP} = 100$~GeV 
are roughly equal to those of a massless LSP\@.  
Raising the LSP mass to $m_{\rm LSP} = 200$~GeV, 
the squark mass limit drops from $680$~GeV to $651$~GeV, 
and we find there is \emph{no} limit for $m_{\rm LSP} = 300$~GeV\@. 
This is consistent with the ``T2'' simplified model studied by
CMS \cite{alphat_supplementalplots}.

There are many other search strategies for supersymmetry that may be
sensitive to more specific models of supersymmetry that have Dirac gaugino
masses.  One of the often-touted searches for supersymmetry
are same-sign dilepton searches, since in the MSSM it is 
straightforward to obtain a significant same-sign dilepton
signature resulting from the chirality-flip of a gaugino due to its
Majorana mass.  In scenarios with Dirac gaugino masses, this
source of same-sign dileptons is completely absent.
Depending on the implementation of the Higgsino sector, 
models with Dirac gaugino masses may or may not have effectively
small Majorana masses and therefore a suppressed same-sign
dilepton signal.  It would certainly be interesting to follow up
on the SSSM with another simplified model of the electroweak
gaugino sector and determine the relative weakness of the 
LHC bounds.  We leave this to future work.

\section*{Acknowledgments}

We thank Ricky Fok, Patrick Fox, and Joe Lykken for 
many valuable conversations. 
GDK was supported in part by a Ben Lee Fellowship from Fermilab and 
in part by the US Department of Energy under contract number 
DE-FG02-96ER40969. AM and GK are supported by Fermilab operated by 
Fermi Research Alliance, LLC under contract number 
DE-AC02-07CH11359 with the US Department of Energy.

\appendix

\renewcommand{\theequation}{\thesection.\arabic{equation}}

\setcounter{figure}{0}
\renewcommand{\thefigure}{\thesection.\arabic{figure}}

\section{Supersoft Supersymmetric Simplified Model (SSSM)}
\label{app:MRSSMlag}

The supersoft supersymmetric simplified model Lagrangian we are 
considering can be expressed as:
\begin{equation}
\mathcal L = \mathcal L_{kin} + \mathcal L_{yuk} + \mathcal L_{decay},
\end{equation}
where $\mathcal L_{kin}$ contains the usual squark and gluino kinetic 
terms, gauge interactions and masses, $\mathcal L_{yuk}$ contains the 
gluino-squark-quark interactions, and $\mathcal L_{decay}$ contains the 
squark-quark-gravitino interactions through which the squarks decay. The 
kinetic term for the squarks is unchanged from the MSSM, while the gluino 
is slightly modified to account for the Dirac character of its mass:
\begin{equation}
\mathcal L_{kin} \supset i\,\bar{\lambda}^a 
\gamma^{\mu}\,(\partial_{\mu}\delta_{ac} + g_s f^{abc} G^b_{\mu} ) 
\lambda_c - \Mg \bar{\lambda}^a\,\lambda_a,
\end{equation}
where $\lambda_a$ is a four-component Dirac bi-spinor. Schematically
\begin{equation}
\lambda_a = \left(\begin{array}{c}  \phi_a \\ \tilde{g}^{*}_a \end{array} 
\right),
\end{equation}
where $\tilde{g}_a$ is usual gluino, in the sense that it is the 
superpartner of the gluon, and $\phi_a$ is the fermionic component of a 
color-adjoint superfield introduced to get a mass with $\tilde{g}_a$.

The Yukawa terms are the same as in the MSSM, however if we want to write 
them in terms of four-component spinors we need to be careful since matter 
(squarks and quarks) couples only to $\tilde{g}_a$ and not to $\phi_a$:
\begin{align}
&\mathcal L_{yuk} = -\sqrt 2 g_s\, (\tilde{u}^*_{L,i} \,t^a 
\bar{\lambda}_a\, P_L\,u_i + \tilde{d}^*_{L,i}\,t^a \bar{\lambda}_a\, P_L\, 
d_i - \nonumber \\
&~~~~~~~~~~~~~~~~ \tilde{u}^*_{R,i}\,t^a \bar{\lambda^c}_a\, P_R\, u_i - 
\tilde{d}^*_{R,i}\,t^a \bar{\lambda^c}_a\, P_R\, d_i) + h.c.,
\end{align}
where $t^a$ are the $SU(3)$ generators, $P_{L,R}$ are the usual chiral 
projection matrices and $i$ labels the flavor index.
 
The gravitino interactions in the SSSM are exactly the same as 
in the MSSM\@.  Approximating interactions with the gravitino by interactions 
with its goldstino longitudinal component, we have
\begin{equation}
\mathcal L_{decay} = \frac{i}{\sqrt 3\,M_P\, m_{3/2} } \bar q_{\omega, 
i}\,\gamma^{\mu}\, P_{\omega} \gamma^{\nu}\, D_{\nu} \tilde{q}_{\omega,i}\, 
(\partial_{\mu} \tilde{G}) + h. c.
\end{equation}
for quark $q$ with helicity $\omega$ and flavor $i$.
 
In practice, the SSSM contains only two free parameters, 
the mass of the (Dirac) gluino and the common mass for the first and 
second generation squarks (both left- and right-handed). 
The gravitino interaction parameters are irrelevant as we assume 
the branching fraction of squark to quark plus gravitino to be 100\%
and the decay is prompt.  The Lorentz form of the 
interactions is important as it determines the kinematics of the final jets, 
which in turn sets the acceptance for a given analysis.

\section{Acceptances for the Analyses}
\label{app:acceptance}

\begin{figure}[!t]
\includegraphics[width=0.45\textwidth]{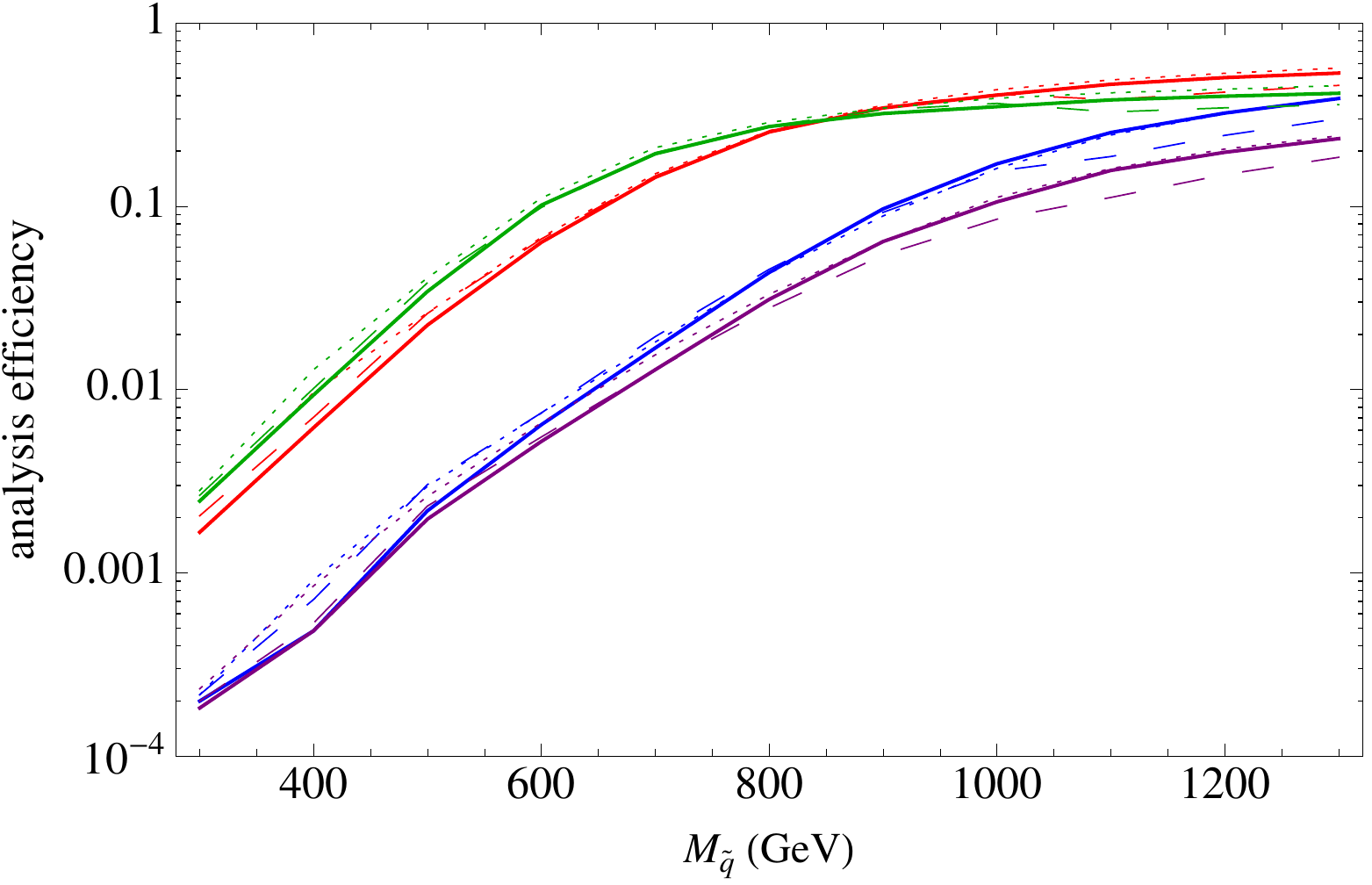}
\includegraphics[width=0.45\textwidth]{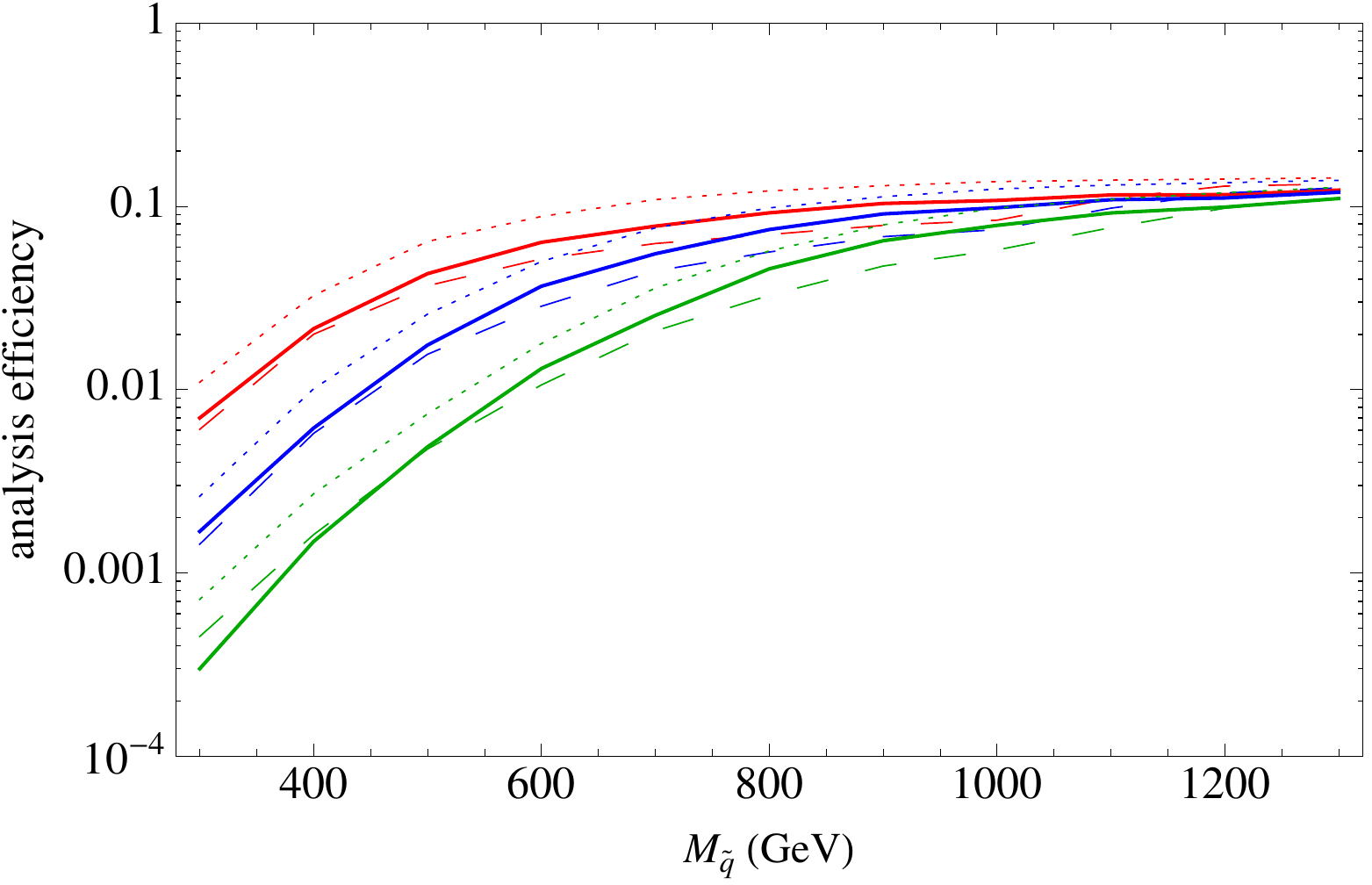}
\includegraphics[width=0.45\textwidth]{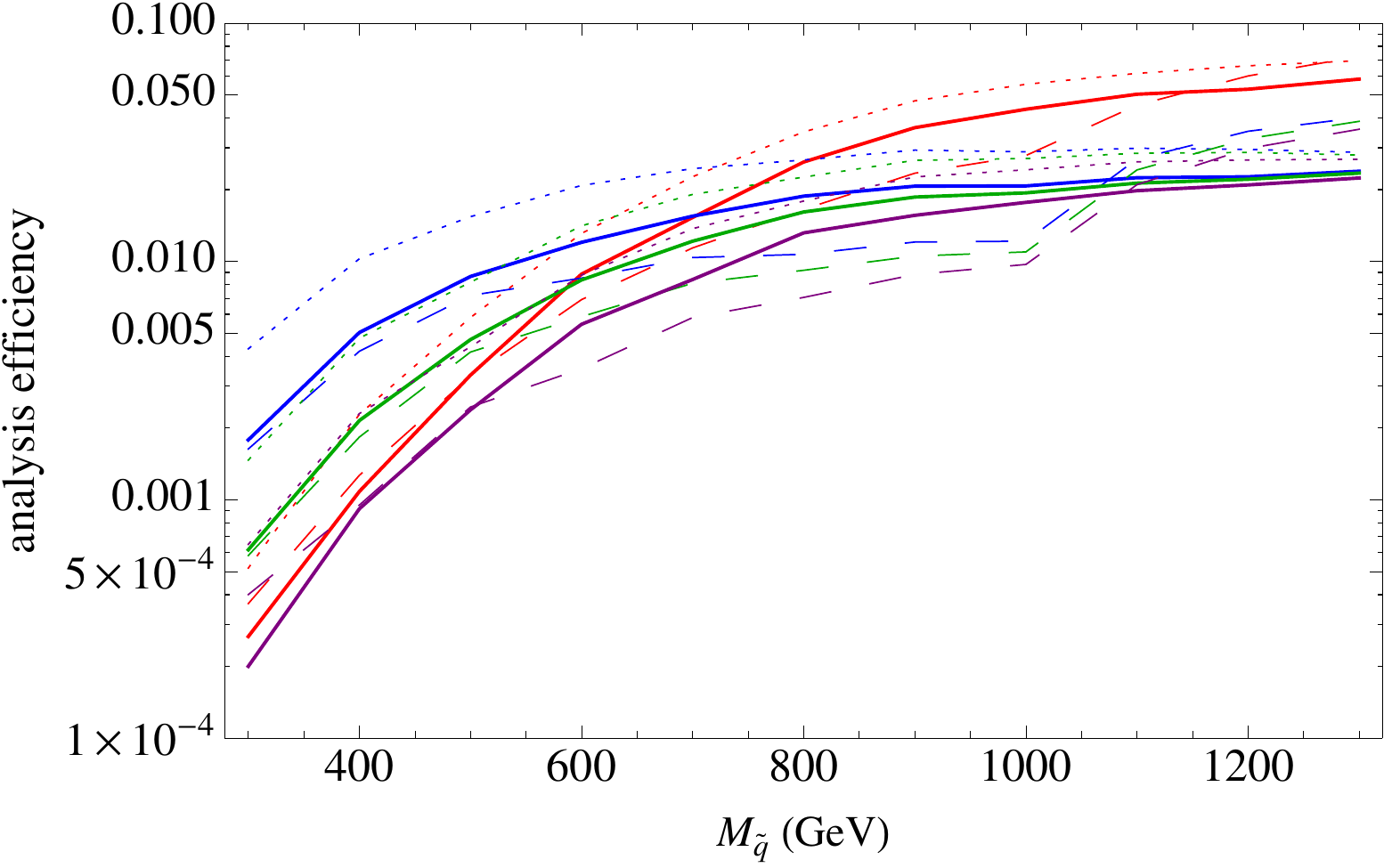}
\caption{Acceptance times efficiency for the eleven channels of the 
ATLAS analysis. The top panel shows $A\cdot\epsilon$ for the 
lower multiplicity channels:  
2j (A) medium in red,
2j (A) tight in blue, 
2j (A$'$) medium in green, and 
3j (B) tight in purple. 
The middle multiplicity (4j) channels are shown in the middle panel: 
loose $m_{eff}(inc.)$ in red, medium in blue, tight in green. 
Finally, the highest multiplicity channels are shown in the bottom panel: 
5j (D) tight in red, 
6j (E) loose in blue, 
6j (E) medium in green, and 
6j (E) tight in purple. 
In all panels, the solid lines correspond to the acceptance times 
efficiency within the SSSM, 
the dotted lines correspond to the ``equal MSSM'' simplified model
with $\tilde{M}_3 = \Msq$,
and the dashed lines correspond to the 
``heavy MSSM'' simplified model with $\tilde{M}_3 = 5$~TeV\@.}
\label{fig:atlas_limits}
\end{figure}

\begin{figure}[!t]
\includegraphics[width=0.45\textwidth]{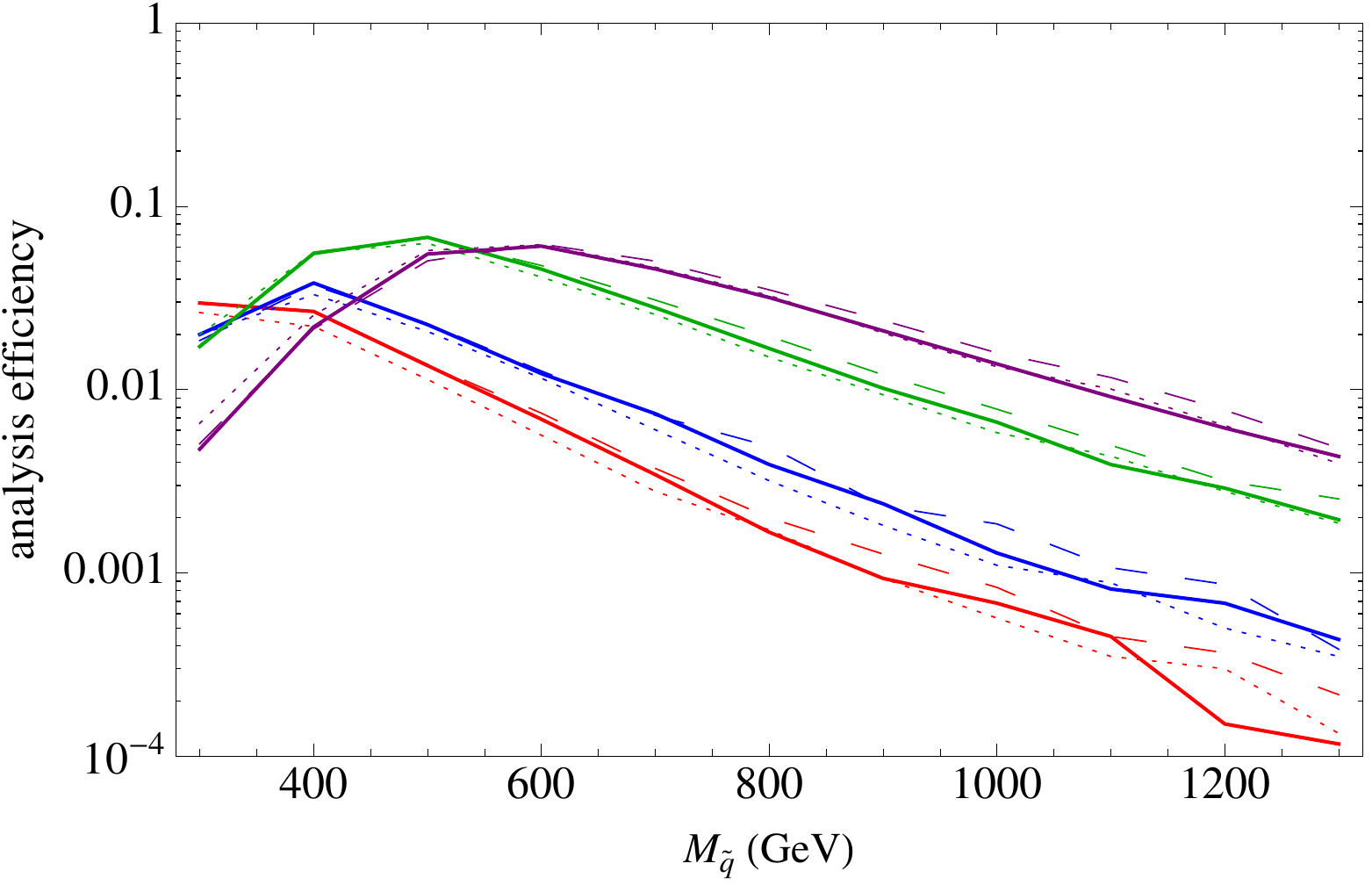}
\includegraphics[width=0.45\textwidth]{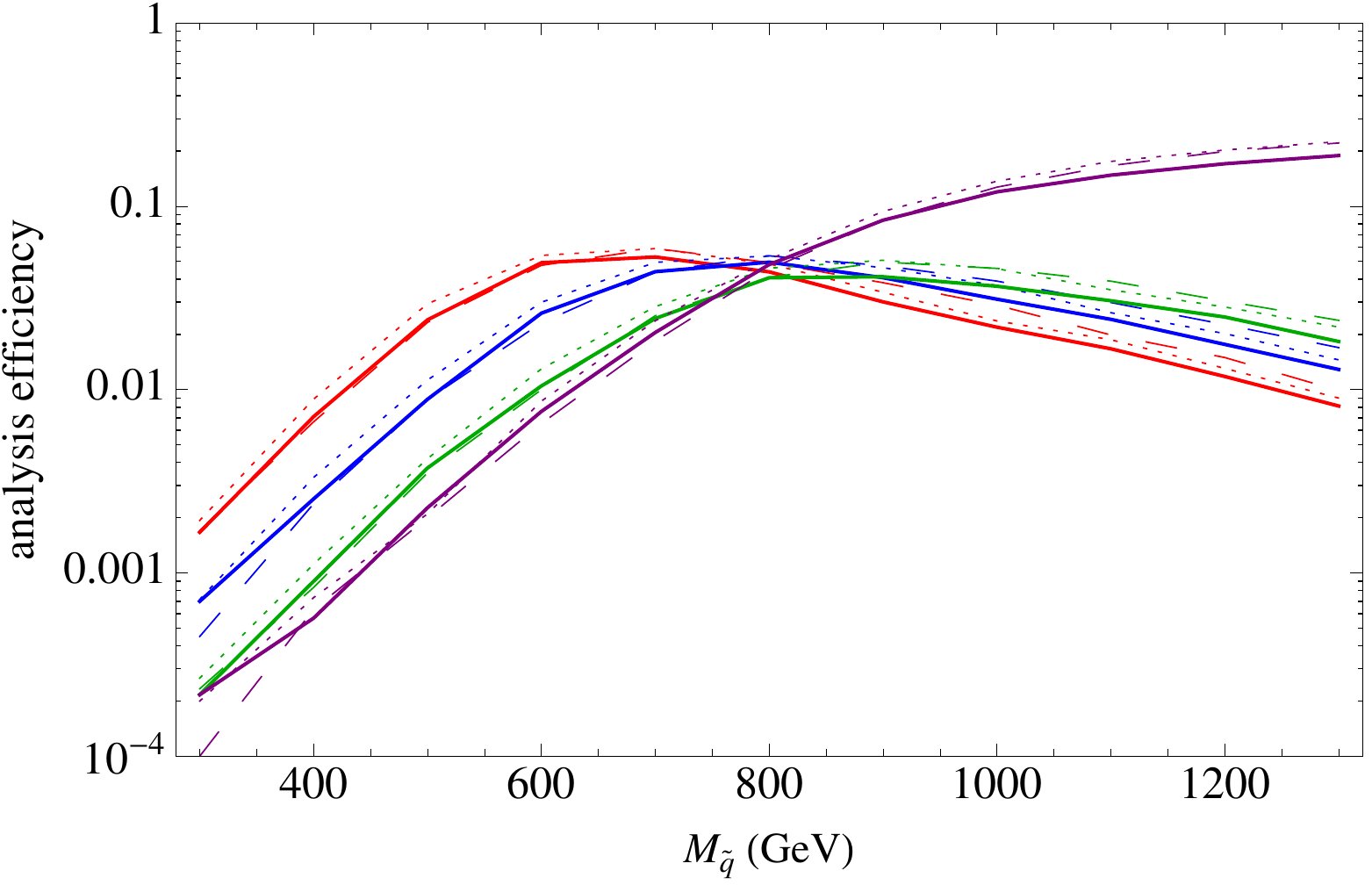}
\caption{Acceptance times efficiency for the 
SSSM (solid), 
``equal MSSM'' simplified model ($\tilde{M}_3 = \Msq$) (dotted), and 
``heavy MSSM'' simplified model ($\tilde{M}_3 = 5$~TeV) (dashed)
models using the CMS $\alpha_T$ analysis. 
The color indicates the which $H_T$ bin was used. In the top panel, 
red shows $H_T = 275$-$325$~GeV, 
blue shows $325$-$375$~GeV, 
green for $375$-$475$~GeV and 
purple for $475$-$575$~GeV\@. 
Similarly, in the bottom panel 
red shows $575$-$675$~GeV, 
blue is $675$-$775$~GeV, 
green is $775$-$875$~GeV and 
purple is $> 875$~GeV\@.}
\label{fig:cms_limits}
\end{figure}

\begin{figure}[!h]
\includegraphics[width=0.45\textwidth]{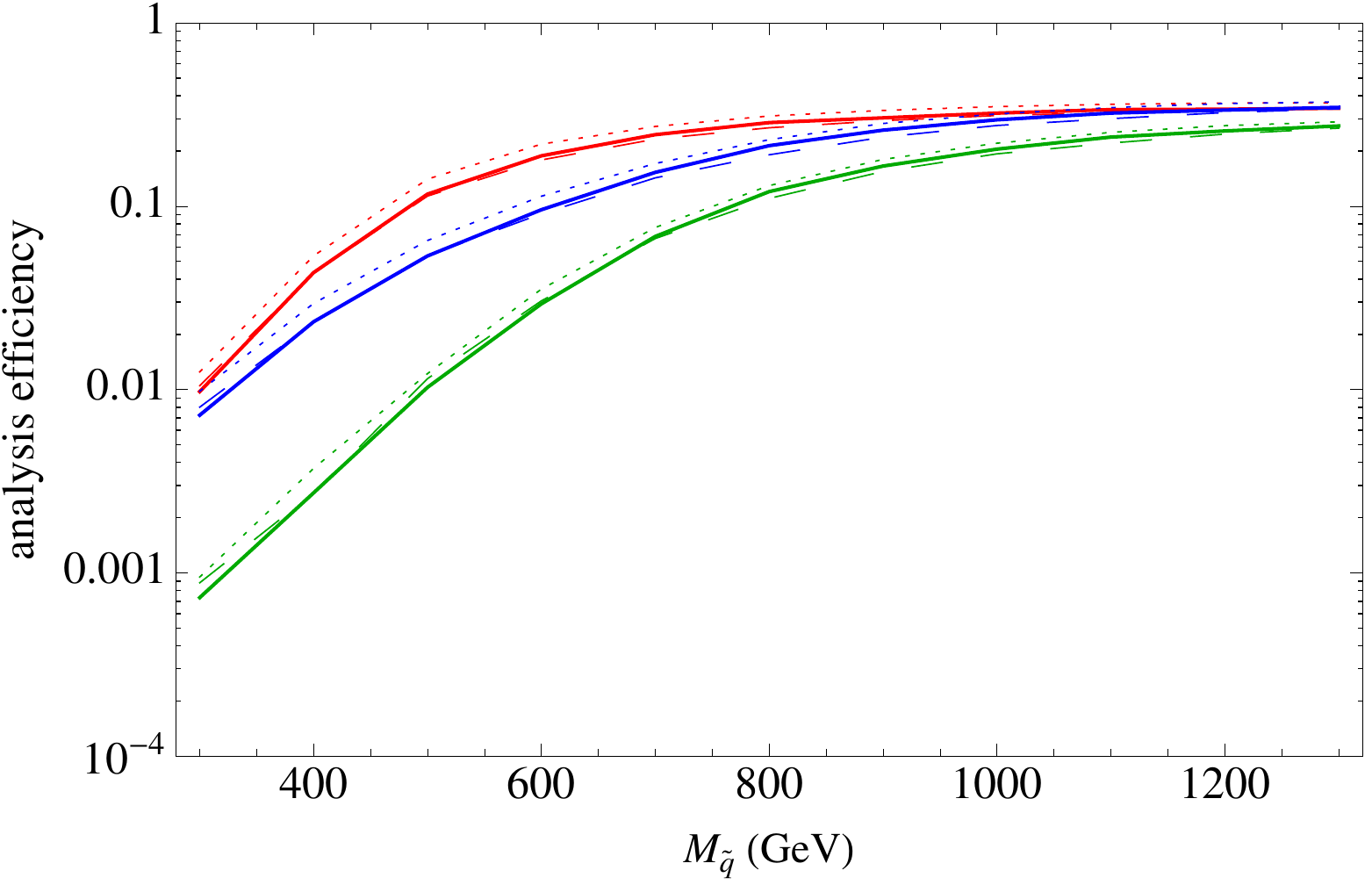}
\caption{Acceptance $\times$ efficiency factors for the CMS search based 
on $H_T, \slashchar{E}_T$. The line hatching follows the same convention as 
Fig.~\ref{fig:cms_limits}. 
Red shows the limits from baseline selection 
plus $\slashchar{E}_T > 350$~GeV, $H_T > 500$~GeV, 
blue is baseline plus $H_T > 800$~GeV and green shows baseline + 
$\slashchar{E}_T > 500$~GeV, $H_T > 800$~GeV\@.}
\label{fig:cms_MHT}
\end{figure}

\begin{figure}[!h]
\includegraphics[width=0.45\textwidth]{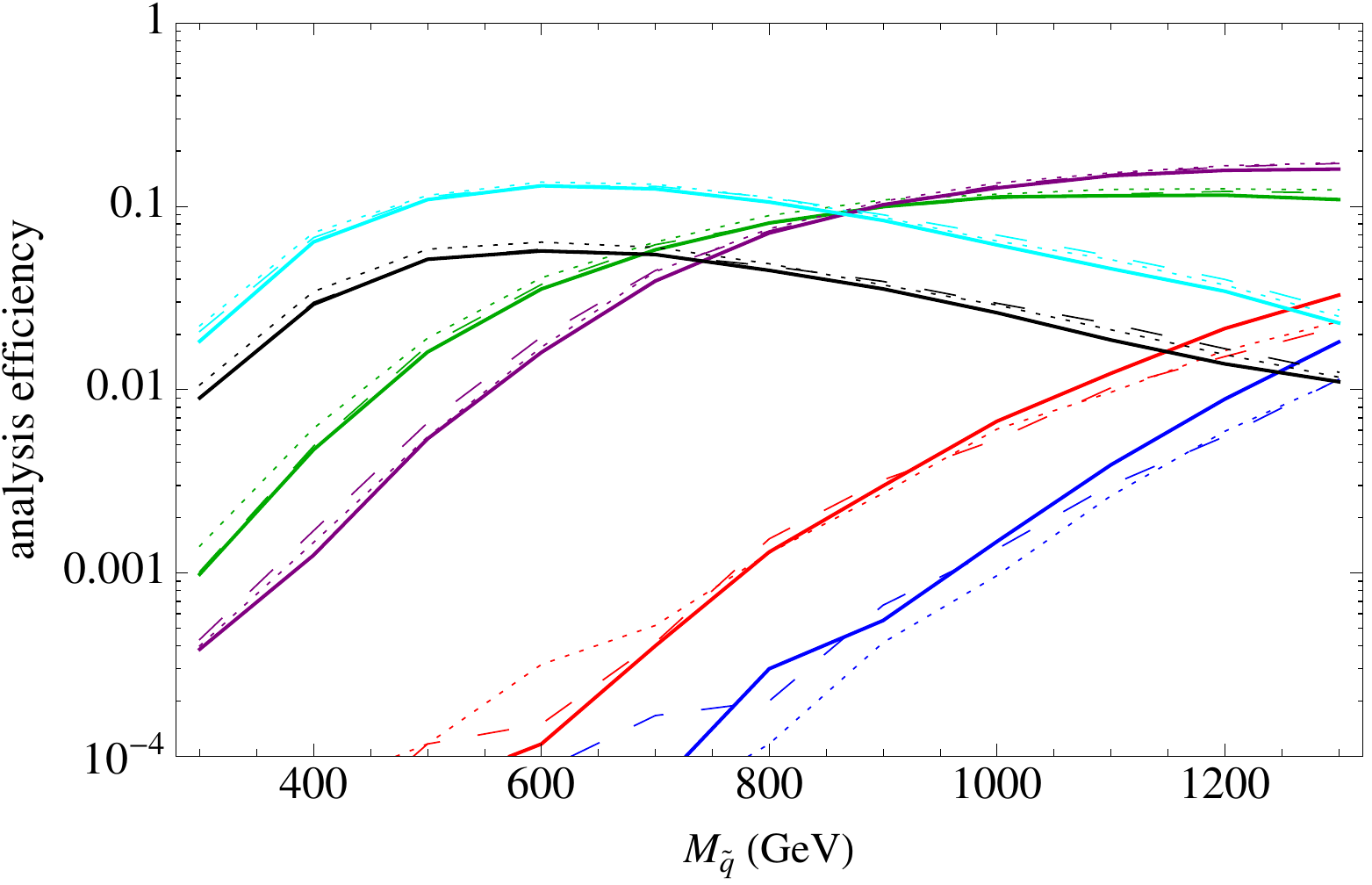}
\caption{Acceptance times efficiency for the CMS search based on the 
hadronic channel of the \emph{razor} analysis.  
The line hatching follows the same convention as 
Fig.~\ref{fig:cms_limits}. 
The six groups of lines correspond to analyses regions 
S1 (red), S2 (blue), S3 (green), S4 (purple), S4 (black) and S6 (cyan).}
\label{fig:cms_razor}
\end{figure}

In this Appendix we collect the series of Figures showing the acceptances
for the various analyses discussed in detail in Sec.~\ref{sec:lhclimits}.
Fig.~\ref{fig:atlas_limits} 
shows the acceptance for 
the ATLAS jets plus missing energy search described in Sec.~\ref{sec:ATLlim}; 
Fig.~\ref{fig:cms_limits}
shows the acceptance for 
the CMS $\alpha_T$ search described in Sec.~\ref{ss:alphat};
Fig.~\ref{fig:cms_MHT}
shows the acceptance for 
the CMS jets plus missing energy search described in 
Sec.~\ref{ss:cmsjetsmissing};
Fig.~\ref{fig:cms_razor}
shows the acceptance for 
the CMS \emph{razor} search described in Sec.~\ref{ss:razor}.

\cleardoublepage 

\begin{thebibliography}{99}

\bibitem{Dimopoulos:1995mi} 
  S.~Dimopoulos and G.~F.~Giudice,
  Phys.\ Lett.\ B {\bf 357}, 573 (1995)
  [hep-ph/9507282].

\bibitem{Cohen:1996vb}
  A.~G.~Cohen, D.~B.~Kaplan, A.~E.~Nelson,
  Phys.\ Lett.\  {\bf B388}, 588-598 (1996).
  [hep-ph/9607394].

\bibitem{ArkaniHamed:2004yi}
  N.~Arkani-Hamed, S.~Dimopoulos, G.~F.~Giudice, A.~Romanino,
  Nucl.\ Phys.\  {\bf B709}, 3-46 (2005).
  [arXiv:hep-ph/0409232 [hep-ph]].




\bibitem{Essig:2011qg} 
  R.~Essig, E.~Izaguirre, J.~Kaplan and J.~G.~Wacker,
  JHEP {\bf 1201}, 074 (2012)
  [arXiv:1110.6443 [hep-ph]].

\bibitem{Kats:2011qh} 
  Y.~Kats, P.~Meade, M.~Reece and D.~Shih,
  arXiv:1110.6444 [hep-ph].

\bibitem{Brust:2011tb} 
  C.~Brust, A.~Katz, S.~Lawrence and R.~Sundrum,
  arXiv:1110.6670 [hep-ph].

\bibitem{Papucci:2011wy} 
  M.~Papucci, J.~T.~Ruderman and A.~Weiler,
  arXiv:1110.6926 [hep-ph].

\bibitem{Feng:2011aa} 
  J.~L.~Feng, K.~T.~Matchev and D.~Sanford,
  arXiv:1112.3021 [hep-ph].

\bibitem{Csaki:2012fh} 
  C.~Csaki, L.~Randall and J.~Terning,
  arXiv:1201.1293 [hep-ph].

\bibitem{Larsen:2012rq} 
  G.~Larsen, Y.~Nomura and H.~L.~L.~Roberts,
  arXiv:1202.6339 [hep-ph].





\bibitem{Barbier:2004ez}
  R.~Barbier {\it et al.},
  Phys.\ Rept.\  {\bf 420}, 1 (2005)
  [arXiv:hep-ph/0406039].

\bibitem{Carpenter:2007zz} 
  L.~M.~Carpenter, D.~E.~Kaplan and E.~-J.~Rhee,
  Phys.\ Rev.\ Lett.\  {\bf 99}, 211801 (2007)
  [hep-ph/0607204].

\bibitem{Csaki:2011ge} 
  C.~Csaki, Y.~Grossman and B.~Heidenreich,
  arXiv:1111.1239 [hep-ph].

\bibitem{Dreiner:2012mn} 
  H.~K.~Dreiner and T.~Stefaniak,
  arXiv:1201.5014 [hep-ph].

\bibitem{Allanach:2012vj} 
  B.~C.~Allanach and B.~Gripaios,
  arXiv:1202.6616 [hep-ph].

\bibitem{compressed}
  T.~J.~LeCompte and S.~P.~Martin,
  Phys.\ Rev.\ D {\bf 84}, 015004 (2011)
  [arXiv:1105.4304 [hep-ph]];
  T.~J.~LeCompte and S.~P.~Martin,
  arXiv:1111.6897 [hep-ph].

\bibitem{stealth}
  J.~Fan, M.~Reece and J.~T.~Ruderman,
  JHEP {\bf 1111}, 012 (2011)
  [arXiv:1105.5135 [hep-ph]];
  J.~Fan, M.~Reece and J.~T.~Ruderman,
  arXiv:1201.4875 [hep-ph].



\bibitem{Fayet:1978qc} 
  P.~Fayet,
  Phys.\ Lett.\ B {\bf 78}, 417 (1978).

\bibitem{Polchinski:1982an} 
  J.~Polchinski and L.~Susskind,
  Phys.\ Rev.\ D {\bf 26}, 3661 (1982).

\bibitem{Hall:1990hq} 
  L.~J.~Hall and L.~Randall,
  Nucl.\ Phys.\ B {\bf 352}, 289 (1991).

\bibitem{Fox:2002bu}
  P.~J.~Fox, A.~E.~Nelson, N.~Weiner,
  JHEP {\bf 0208}, 035 (2002).
  [hep-ph/0206096].



\bibitem{Alwall:2008ag}
  J.~Alwall, P.~Schuster, N.~Toro,
  Phys.\ Rev.\  {\bf D79}, 075020 (2009).
  [arXiv:0810.3921 [hep-ph]].

\bibitem{Alves:2011wf}
  D.~Alves, N.~Arkani-Hamed, S.~Arora, Y.~Bai, M.~Baumgart, J.~Berger, M.~Buckley, B.~Butler {\it et al.},
  [arXiv:1105.2838 [hep-ph]].

\bibitem{Nelson:2002ca} 
  A.~E.~Nelson, N.~Rius, V.~Sanz and M.~Unsal,
  JHEP {\bf 0208}, 039 (2002)
  [hep-ph/0206102].

\bibitem{Chacko:2004mi}
  Z.~Chacko, P.~J.~Fox, H.~Murayama,
  Nucl.\ Phys.\  {\bf B706}, 53-70 (2005).
  [hep-ph/0406142].

\bibitem{Carpenter:2005tz} 
  L.~M.~Carpenter, P.~J.~Fox and D.~E.~Kaplan,
  hep-ph/0503093.

\bibitem{Antoniadis:2005em} 
  I.~Antoniadis, A.~Delgado, K.~Benakli, M.~Quiros and M.~Tuckmantel,
  Phys.\ Lett.\ B {\bf 634}, 302 (2006)
  [hep-ph/0507192].

\bibitem{Nomura:2005rj} 
  Y.~Nomura, D.~Poland and B.~Tweedie,
  Nucl.\ Phys.\ B {\bf 745}, 29 (2006)
  [hep-ph/0509243].

\bibitem{Antoniadis:2006uj} 
  I.~Antoniadis, K.~Benakli, A.~Delgado and M.~Quiros,
  Adv.\ Stud.\ Theor.\ Phys.\  {\bf 2}, 645 (2008)
  [hep-ph/0610265].

\bibitem{Kribs:2007ac}
  G.~D.~Kribs, E.~Poppitz, N.~Weiner,
  Phys.\ Rev.\  {\bf D78}, 055010 (2008).
  [arXiv:0712.2039 [hep-ph]].

\bibitem{Amigo:2008rc} 
  S.~D.~L.~Amigo, A.~E.~Blechman, P.~J.~Fox and E.~Poppitz,
  JHEP {\bf 0901}, 018 (2009)
  [arXiv:0809.1112 [hep-ph]].

\bibitem{Benakli:2008pg} 
  K.~Benakli and M.~D.~Goodsell,
  Nucl.\ Phys.\ B {\bf 816}, 185 (2009)
  [arXiv:0811.4409 [hep-ph]].

\bibitem{Blechman:2009if} 
  A.~E.~Blechman,
  Mod.\ Phys.\ Lett.\ A {\bf 24}, 633 (2009)
  [arXiv:0903.2822 [hep-ph]].

\bibitem{Carpenter:2010as} 
  L.~M.~Carpenter,
  arXiv:1007.0017 [hep-th].

\bibitem{Kribs:2010md} 
  G.~D.~Kribs, T.~Okui and T.~S.~Roy,
  Phys.\ Rev.\ D {\bf 82}, 115010 (2010)
  [arXiv:1008.1798 [hep-ph]].

\bibitem{Abel:2011dc} 
  S.~Abel and M.~Goodsell,
  JHEP {\bf 1106}, 064 (2011)
  [arXiv:1102.0014 [hep-th]].

\bibitem{Frugiuele:2011mh} 
  C.~Frugiuele and T.~Gregoire,
  Phys.\ Rev.\ D {\bf 85}, 015016 (2012)
  [arXiv:1107.4634 [hep-ph]].

\bibitem{Itoyama:2011zi} 
  H.~Itoyama and N.~Maru,
  arXiv:1109.2276 [hep-ph].



















\bibitem{Hisano:2006mv} 
  J.~Hisano, M.~Nagai, T.~Naganawa and M.~Senami,
  Phys.\ Lett.\ B {\bf 644}, 256 (2007)
  [hep-ph/0610383].

\bibitem{Hsieh:2007wq} 
  K.~Hsieh,
  Phys.\ Rev.\ D {\bf 77}, 015004 (2008)
  [arXiv:0708.3970 [hep-ph]].

\bibitem{Blechman:2008gu} 
  A.~E.~Blechman and S.~-P.~Ng,
  JHEP {\bf 0806}, 043 (2008)
  [arXiv:0803.3811 [hep-ph]].

\bibitem{Kribs:2008hq} 
  G.~D.~Kribs, A.~Martin and T.~S.~Roy,
  JHEP {\bf 0901}, 023 (2009)
  [arXiv:0807.4936 [hep-ph]].

\bibitem{Choi:2008pi} 
  S.~Y.~Choi, M.~Drees, A.~Freitas and P.~M.~Zerwas,
  Phys.\ Rev.\ D {\bf 78}, 095007 (2008)
  [arXiv:0808.2410 [hep-ph]].

\bibitem{Plehn:2008ae} 
  T.~Plehn and T.~M.~P.~Tait,
  J.\ Phys.\ G G {\bf 36}, 075001 (2009)
  [arXiv:0810.3919 [hep-ph]].

\bibitem{Harnik:2008uu} 
  R.~Harnik and G.~D.~Kribs,
  Phys.\ Rev.\ D {\bf 79}, 095007 (2009)
  [arXiv:0810.5557 [hep-ph]].

\bibitem{Choi:2008ub} 
  S.~Y.~Choi, M.~Drees, J.~Kalinowski, J.~M.~Kim, E.~Popenda and P.~M.~Zerwas,
  Phys.\ Lett.\ B {\bf 672}, 246 (2009)
  [arXiv:0812.3586 [hep-ph]].

\bibitem{Kribs:2009zy} 
  G.~D.~Kribs, A.~Martin and T.~S.~Roy,
  JHEP {\bf 0906}, 042 (2009)
  [arXiv:0901.4105 [hep-ph]].

\bibitem{Belanger:2009wf} 
  G.~Belanger, K.~Benakli, M.~Goodsell, C.~Moura and A.~Pukhov,
  JCAP {\bf 0908}, 027 (2009)
  [arXiv:0905.1043 [hep-ph]].

\bibitem{Benakli:2009mk} 
  K.~Benakli and M.~D.~Goodsell,
  Nucl.\ Phys.\ B {\bf 830}, 315 (2010)
  [arXiv:0909.0017 [hep-ph]].

\bibitem{Kumar:2009sf} 
  A.~Kumar, D.~Tucker-Smith and N.~Weiner,
  JHEP {\bf 1009}, 111 (2010)
  [arXiv:0910.2475 [hep-ph]].

\bibitem{Chun:2009zx} 
  E.~J.~Chun, J.~-C.~Park and S.~Scopel,
  JCAP {\bf 1002}, 015 (2010)
  [arXiv:0911.5273 [hep-ph]].

\bibitem{Benakli:2010gi} 
  K.~Benakli and M.~D.~Goodsell,
  Nucl.\ Phys.\ B {\bf 840}, 1 (2010)
  [arXiv:1003.4957 [hep-ph]].

\bibitem{Fok:2010vk} 
  R.~Fok and G.~D.~Kribs,
  Phys.\ Rev.\ D {\bf 82}, 035010 (2010)
  [arXiv:1004.0556 [hep-ph]].

\bibitem{DeSimone:2010tf} 
  A.~De Simone, V.~Sanz and H.~P.~Sato,
  Phys.\ Rev.\ Lett.\  {\bf 105}, 121802 (2010)
  [arXiv:1004.1567 [hep-ph]].

\bibitem{Choi:2010gc} 
  S.~Y.~Choi, D.~Choudhury, A.~Freitas, J.~Kalinowski, J.~M.~Kim and P.~M.~Zerwas,
  JHEP {\bf 1008}, 025 (2010)
  [arXiv:1005.0818 [hep-ph]].

\bibitem{Choi:2010an} 
  S.~Y.~Choi, D.~Choudhury, A.~Freitas, J.~Kalinowski and P.~M.~Zerwas,
  Phys.\ Lett.\ B {\bf 697}, 215 (2011)
  [Erratum-ibid.\ B {\bf 698}, 457 (2011)]
  [arXiv:1012.2688 [hep-ph]].

\bibitem{Benakli:2011kz} 
  K.~Benakli, M.~D.~Goodsell and A.~-K.~Maier,
  Nucl.\ Phys.\ B {\bf 851}, 445 (2011)
  [arXiv:1104.2695 [hep-ph]].

\bibitem{Kumar:2011np} 
  P.~Kumar and E.~Ponton,
  JHEP {\bf 1111}, 037 (2011)
  [arXiv:1107.1719 [hep-ph]].

\bibitem{Heikinheimo:2011fk} 
  M.~Heikinheimo, M.~Kellerstein and V.~Sanz,
  arXiv:1111.4322 [hep-ph].

\bibitem{Fuks:2012im} 
  B.~Fuks,
  arXiv:1202.4769 [hep-ph].

\bibitem{atlas_jetsmet}
  G.~Aad {\it et al.}  [ATLAS Collaboration],
  ATLAS-CONF-2011-155,
  arXiv:1109.6572 [hep-ex]; 
  G.~Aad {\it et al.}  [ATLAS Collaboration],
  ATLAS-CONF-2012-033
  
\bibitem{cms_alpha_t} 
  S.~Chatrchyan {\it et al.}  [CMS Collaboration],
  Phys.\ Rev.\ Lett.\  {\bf 107}, 221804 (2011)
  [arXiv:1109.2352 [hep-ex]].

\bibitem{cms_mht}
  CMS Collaboration, CMS-PAS-SUS-11-004.

\bibitem{razor}
  S.~Chatrchyan {\it et al.}  [CMS Collaboration],
  Phys.\ Rev.\ D {\bf 85}, 012004 (2012)
  [arXiv:1107.1279 [hep-ex]];
  CMS Collaboration, CMS-PAS-SUS-12-005.

\bibitem{Sjostrand:2006za} 
  T.~Sjostrand, S.~Mrenna and P.~Z.~Skands,
  JHEP {\bf 0605}, 026 (2006)
  [hep-ph/0603175].

\bibitem{Ovyn:2009tx} 
  S.~Ovyn, X.~Rouby and V.~Lemaitre,
  arXiv:0903.2225 [hep-ph].

\bibitem{alphat_supplementalplots}  
  https://twiki.cern.ch/twiki/bin/view/CMSPublic\\ 
  ~~~~~~~~~~~~~  /PhysicsResultsSUS11003

\bibitem{prospino}
  W.~Beenakker, R.~Hopker, M.~Spira and P.~M.~Zerwas,
  Nucl.\ Phys.\ B {\bf 492}, 51 (1997)
  [hep-ph/9610490];
  W.~Beenakker, M.~Kramer, T.~Plehn, M.~Spira and P.~M.~Zerwas,
  Nucl.\ Phys.\ B {\bf 515}, 3 (1998)
  [hep-ph/9710451];
  W.~Beenakker, M.~Klasen, M.~Kramer, T.~Plehn, M.~Spira and P.~M.~Zerwas,
  Phys.\ Rev.\ Lett.\  {\bf 83}, 3780 (1999)
  [Erratum-ibid.\  {\bf 100}, 029901 (2008)]
  [hep-ph/9906298];
  M.~Spira,
  hep-ph/0211145;
  T.~Plehn,
  Czech.\ J.\ Phys.\  {\bf 55}, B213 (2005)
  [hep-ph/0410063].

\bibitem{conway1}
J.~Conway, K.~Maeshima, \\ CDF/PUB/EXOTIC/PUBLIC/4476.

\bibitem{conway}
  J.~Conway, CDF/PUB/STATISTICS/PUBLIC/6428.

\bibitem{Cranmer:2010hk} 
  K.~Cranmer and I.~Yavin,
  JHEP {\bf 1104}, 038 (2011)
  [arXiv:1010.2506 [hep-ex]].

\bibitem{Aad:2011qa}
  G.~Aad {\it et al.}  [Atlas Collaboration],
  arXiv:1110.2299 [hep-ex]., 
  G.~Aad {\it et al.}  [Atlas Collaboration]
ATLAS-CONF-2012-037
 
\bibitem{Rogan:2010kb}
  C.~Rogan,
  [arXiv:1006.2727 [hep-ph]].

\end{thebibliography}
\end{document}